\documentclass[fleqn,
superscriptaddress,
preprintnumbers,
nofootinbib,
amsmath,amssymb,
aps,
prd,
twocolumn,
]{revtex4-1}

\usepackage[T1]{fontenc}
\usepackage{ae,aecompl}

\usepackage{graphicx}	
\usepackage{amsmath}	
\usepackage{amssymb}	
\usepackage{xspace}
\usepackage{xcolor}
\usepackage{hyperref}

\setcitestyle{notesep={}}

\newcommand{\dcc}{LIGO-P1800394-v4\xspace}

\newcommand{\erfc}{\mathrm{erfc}}

\newcommand{\fgw}{f_\mathrm{gw}}

\newcommand{\updated}[1]{#1} 
\newcommand{\newupdated}[1]{#1} 
\newcommand{\gw}[1][]{gravitational wave#1 (GW#1)\renewcommand{\gw}[1][]{GW##1\xspace}\xspace}
\newcommand{\cw}[1][]{continuous wave#1 (CW#1)\renewcommand{\cw}[1][]{CW##1\xspace}\xspace}
\newcommand{\ns}[1][]{neutron star#1 (NS#1)\renewcommand{\ns}[1][]{NS##1\xspace}\xspace}
\newcommand{\bns}[1][]{binary neutron star#1 (BNS#1)\renewcommand{\bns}[1][]{BNS##1\xspace}\xspace}
\newcommand{\taylor}{\mathcal{T}}

\begin{document}

\title{The Adaptive Transient Hough method for long-duration gravitational wave transients}

\newcommand{\uib}{Departament de F\'isica, Universitat de les Illes Balears, IAC3--IEEC, Cra. Valldemossa Km. 7.5, E-07122 Palma de Mallorca, Spain}
\author{Miquel Oliver}
 \email{miquel.oliver@ligo.org}
 \affiliation{\uib}
\author{David Keitel}
 \email{david.keitel@ligo.org}
 \affiliation{University of Portsmouth, Institute of Cosmology and Gravitation, Portsmouth PO1 3FX, United Kingdom}
\author{Alicia M. Sintes}
 \email{alicia.sintes@uib.es}
 \affiliation{\uib}

\date{\today, version: \dcc}

\begin{abstract}
This paper describes a new semi-coherent method to search for transient gravitational waves of intermediate duration (hours to days).
\updated{In order to search for newborn isolated neutron stars
with their possibly very rapid spin-down,
we model the frequency evolution as a power law.}
The search uses short Fourier transforms from the output of ground-based gravitational wave detectors and applies a weighted Hough transform,
\updated{also taking into account the signal's amplitude evolution}.
We present the technical details for implementing the algorithm, its statistical properties, and a sensitivity estimate.
A first example application of this method was in the search for GW170817 post-merger signals,
\updated{and we verify the estimated sensitivity with simulated signals for this case}.
\end{abstract}

\pacs{Valid PACS appear here}
\maketitle



\section{INTRODUCTION}

The advanced \gw detector era has provided us with multiple detections from binary compact objects
\citep{LIGOScientific:2018mvr}
including GW170817,
the first \bns coalescence \citep{GW170817}.
This detection motivated the development of the new search method presented in this paper,
focusing on the possible birth of a rapidly rotating highly magnetized \ns
spinning down through some combination of GW and electromagnetic emission.
For a very massive remnant, the collapse would occur in a short time scale
(as explored in \cite{Abbott:2017dke,Abbott:2018wiz}),
but for low total mass
and some equations of state, the emitted GW signal could have an intermediate duration on the order of hours to days 
\citep{Baiotti:2016qnr,Piro:2017zec}.

This regime of GW signal durations has long been mostly unexplored from the data analysis side.
\updated{The expected rapid frequency and amplitude evolution,
in combination with observation times still much longer than e.g. for individual binary coalescences,
pose unique challenges on analysis algorithms.}
Other pre-existing or recently developed methods to search for intermediate-duration signals include
the Stochastic Transient Analysis Multi-detector Pipeline (STAMP) \citep{Thrane2011},
the \updated{Hidden Markov model} Viterbi algorithm \citep{Sun:2018owi}
and a generalization of the FrequencyHough method \citep{Miller:2018rbg}.
The first two are \updated{generic} unmodeled searches,
while the last is a modeled search \updated{for power-law spin-downs}
similar to the one described in this paper.
Together with those three pipelines,
our new \textbf{Adaptive Transient Hough (ATrHough)} method has already contributed to the search for a long-duration transient signal
from a putative \ns remnant of GW170817 described in \cite{GW170817remnant}.

The Adaptive Transient Hough is a semi-coherent analysis adapted from
the SkyHough \citep{WHough,Sintes2006,Hough} search for \cw signals.
\updated{Like most other \cw searches \citep{Riles:2017evm},}
the original SkyHough assumes a constant intrinsic amplitude and slowly evolving frequency,
\updated{and hence cannot be used to search for transient GWs with rapid frequency and amplitude evolution
(see quantitative comparison in Sec.~\ref{sec:model}),
for which we have now specifically developed the new method.}

\updated{The ATrHough method will also have wider applicability beyond the case of \bns remnants,
as signals with similar durations and evolutionary behaviour are also
possible from young \ns[s] born through the regular supernova channel
\citep{Palomba2001:mag,DallOsso:2008kll,DallOsso:2014hpa,Lasky:2015olc,DallOsso:2018dos},
emitted either by r-mode oscillations \citep{Owen1998,Andersson:2000mf}
or quadrupolar deformations.}

The paper is organized as follows:
section II briefly describes the expected signal from a remnant \ns.
Section III summarizes the general strategy of a hierarchical search and its implementation,
section IV studies its statistical properties,
and section V introduces the threshold and vetoes required for a robust detection strategy.
Finally section VI presents an estimate for the search sensitivity
and section VII presents our conclusions.

\section{THE TRANSIENT SIGNAL MODEL}
\label{sec:model}

The output of a GW detector can be represented by
\begin{equation}
x(t) = n(t) + h(t),
\end{equation}
where $n(t)$ is the detector noise at time $t$, and $h(t)$ is the strain induced by a GW signal:
\updated{\begin{equation}
h(t) = F_+ (\textbf{n}, \psi,t) h_+ (t) + F_{\times} (\textbf{n},\psi,t) h_{\times} (t),
\end{equation}}
where $F_{+,\times}$ are the detector antenna patterns,
\updated{which depend on a unit-vector $\textbf{n}$ corresponding to the sky location of the source
and on the wave polarization angle $\psi$,
and vary with time due to the movement of the detector frames with the Earth.}
For ground-based detectors with perpendicular arms, the expressions for $F_{+,\times}$ are \citep{grav1}:
\updated{
\begin{subequations}
 \label{eq:FplusFcross}
 \begin{eqnarray}
  F_+(\textbf{n},\psi,t)        &=& a(t;\textbf{n}) \cos2\psi + b(t;\textbf{n})\sin2\psi,\\
  F_{\times}(\textbf{n},\psi,t) &=& b(t;\textbf{n}) \cos2\psi - a(t;\textbf{n})\sin2\psi,
 \end{eqnarray}
\end{subequations}
}
where the functions \updated{$a(t;\textbf{n})$ and $b(t;\textbf{n})$} are independent of $\psi$.
\updated{For convenience, we do not explicitly write out the $\textbf{n}$ and $\psi$ dependence from here on.}
Now the waveforms for the two polarizations $h_{+,\times}$ are:
\begin{subequations}
 \label{eq:hplushcross}
 \begin{eqnarray}
  h_+(t) &=& A_+(t) \cos \Phi(t),\\
  h_{\times}(t) &=& A_{\times}(t) \sin \Phi(t),
 \end{eqnarray}
\end{subequations}
where $\Phi(t)$ is the phase evolution of the signal
and \updated{$A_{+,\times}(t)$ are the (time-varying)} amplitude parameters
depending on the orientation $\cos{\iota}$ of the source
and on the strain amplitude evolution $h_0(t)$ as follows: 
\begin{subequations}
 \label{eq:AplusAcross}
 \begin{eqnarray}
  A_{+}(t)      &=& \frac{1}{2}h_0(t) \; (1+\cos^2{\iota}) \,, \\
  A_{\times}(t) &=& h_0(t) \cos{\iota} \,.
 \end{eqnarray}
\end{subequations}
The time evolution of the dimensionless strain amplitude \updated{$h_0(t)$} depends on the emission mechanism;
\updated{if for example it is due to a constant non-axisymmetrical deformation of the source \ns,
but the frequency decays over time,
the amplitude evolves as
\begin{equation}
 \label{ampm2}
 h_0(t)=\frac{4 \pi^2 G}{c^4}\frac{I_{zz} \epsilon}{d} \fgw^2(t),
\end{equation}
where $c$ is the speed of light,
$I_{zz}$ is the z-z component of the star's moment of inertia with the z-axis being its spin axis,
\mbox{$\epsilon := (I_{xx} - I_{yy})/I_{zz}$} is the equatorial ellipticity of the star,
and $d$ is its distance from Earth.
Another mechanism covered by this method is \gw emission from r-mode oscillations,
which are the result of small velocity and density perturbations of the \ns fluid,
causing a time-varying moment of inertia restored throw Coriolis force;
for these, the amplitude evolution is given by
\begin{equation}
 \label{amp-rmodes}
 h_0(t)=\sqrt{\frac{3}{80 \pi}}\frac{G}{c^5}\frac{1}{d}\alpha M R^3 \widetilde{J} f_\mathrm{r}^3(t),
\end{equation}
where $f_\mathrm{r}$ is the rotation frequency of the source,
\mbox{$\widetilde{J} = 0.01635$} is a dimensionless constant,
$M$ is the \ns mass,
$R$ its radius
and $\alpha$ is a dimensionless amplitude described in more detail in \cite{Owen1998}.}

\updated{Independent of the specific emission scenario,
the amplitude evolution $h_0(t)$ can be written in a more general form as:}
\begin{equation}
 \label{amplitude-model}
 h_0(t)= A_\mathrm{m} \fgw^m(t) ,
\end{equation}
where $m$ and $A_\mathrm{m}$ are constants defined by the emission mechanism.

\updated{To characterize the frequency evolution of a newborn \ns
we apply the waveform model from \cite{Lasky2017,Sarin2018},
originating from the general torque equation
\begin{equation}\label{freqevol}
\dot\Omega=-\kappa \, \Omega^n,
\end{equation}
where $\Omega$ and $\dot\Omega$ are the frequency of rotation of the source and its derivative.
(When we focus on \gw emission due to a non-axisymmetrical shape
and do not consider the free precession case \citep{Zimmermann:1979ip,Jones:2000ud},
the frequency of \gw emission is \mbox{$\fgw=\Omega/\pi$}.)
Furthermore, $n$ is called the star's braking index
and $\kappa$ is associated to the spindown timescale:}
\updated{\begin{equation}
\tau = -\frac{\Omega_0^{1-n}}{\kappa (1-n)}.
\end{equation}}

\updated{The solution of Eq.~\eqref{freqevol} for arbitrary braking index $n$ characterizes the frequency evolution:}
\begin{equation}
\label{model}
\hat{f}_\mathrm{gw}(t)=
\left\{
	\begin{array}{ll}
		f_{\mathrm{gw},0} \left(\frac{t-T_{0}}{\tau }+1\right)^{\frac{1}{1-n}}  & \mathrm{if}\; t \geq T_{0} \,, \\
		0 & \mathrm{if}\; t < T_{0} \,,
	\end{array}
\right.
\end{equation}
\updated{where $f_{\mathrm{gw},0}$ corresponds to the frequency at the start of the emission (\mbox{$t=T_{0}$});
for simplicity let us set \mbox{$T_{0}=0$\,s}.
A braking index of \mbox{$n=5$} corresponds to pure \gw emission from a non-axisymmetric rotator.
This equation can also be applied to r-modes, for which \mbox{$n\lesssim7$}~\citep{Alford:2012yn,Alford:2014pxa}.}

\updated{The Eq.~\eqref{model} frequency evolution model and resulting amplitude evolution as per Eq.~\eqref{amplitude-model}
is the key difference between our new search method and the SkyHough search~\citep{Hough} for \cw signals,
which instead uses a Taylor expansion for the slowly-evolving frequency of mature \ns[s] and assumes constant intrinsic amplitude.
}

\updated{To demonstrate explicitly that such an expansion is unsuited to search for signals with rapid spindowns,
let us consider that the frequency resolution of a fully-coherent \cw-like search over an observation time
is \mbox{$\delta \fgw = 1/T_\mathrm{obs}$}.
Hence, for a Taylor expansion model $\taylor[\fgw(t),s]$ to order $s$ in $\fgw(t)$,
the requirement is
\mbox{$| \fgw(t) - \taylor[\fgw(t),s] | < 1/T_\mathrm{obs}$}.
Now we see that at least a 16th order expansion is required to track sources with
astrophysically relevant example parameters
(compare~\cite{GW170817remnant})
\mbox{$\fgw(0) = 1000$\,Hz}, \mbox{$\tau = 10000$\,s} and \mbox{$n = 5$}
over \mbox{$T_\mathrm{obs} = 5000$\,s},
making this approach computationally prohibitive.
On the other hand, the search method introduced in the following
uses the exact analytical form with its only three free parameters $(n,f_{\mathrm{gw},0},\tau)$
to create a template grid that ensures complete coverage,
while keeping the analysis computationally feasible.}

As in other semi-coherent searches, this method considers as negligible -- and therefore ignores -- relativistic corrections,
and those due to the time delay between the detector and the solar-system barycenter (SSB).
Therefore only the instantaneous signal frequency in the detector frame needs to be calculated:
\begin{equation}\label{doppler}
\fgw(t)=\hat{f}_\mathrm{gw}(t)\Big(1+\frac{\textbf{v(t)} \cdot \textbf{n}}{c}\Big) \,,
\end{equation}
where $\textbf{v(t)}$ is the detector velocity with respect to the SSB frame.
Note that now the time coordinate $t$ corresponds to time at the detector.

\section{The Adaptive Transient Hough Method}

This section discusses the implementation of the Adaptative Transient Hough (ATrHough) method,
a pipeline based on the semi-coherent SkyHough search for \cw[s] described in \cite{Hough,WHough}.
The common ground of both searches is the use of a weighted Hough transform on Short-time Fourier Transforms (SFTs) as the input data.
The Hough transform is an algorithm widely used in pattern recognition;
here the pattern is defined by the frequency evolution of the signal in the detector data.
In both \cw and transient cases,
the weights take into account the amplitude modulation of the signal, caused by the antenna pattern,
and the changing noise floor between SFTs.
But as a difference to the \cw SkyHough search, the new ATrHough method also includes the source amplitude evolution in the weights.

\updated{Together with the power-law frequency evolution model from Eq.~\eqref{model},
the amplitude weights allow a sensitive search for transient signals from rapidly evolving newborn \ns[s].
Meanwhile, the main framework and statistical properties are the same as in the SkyHough method.
In the following we summarize them in the new context,
and add the required transient-specific details.}

\subsection{Length of Short-duration Fourier Transforms}

\begin{figure}
 \includegraphics[width=\columnwidth]{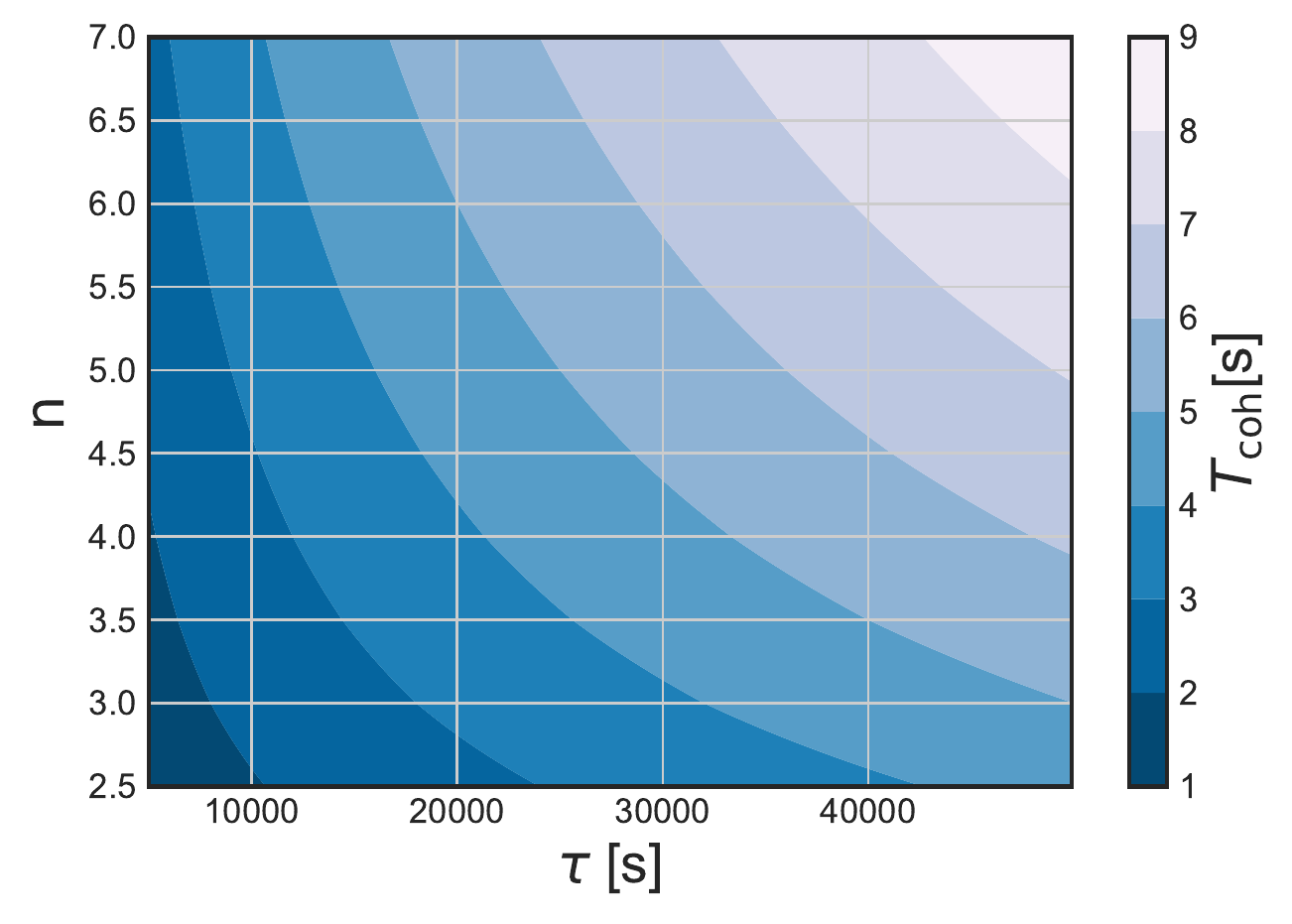}
 \vspace{-\baselineskip}
 \caption{
  \label{fig:tcoh}
  Search setup:
  The maximum coherence length
  \mbox{$T_\mathrm{coh} = \sqrt{(n-1) \tau }/\sqrt{2 f_{\mathrm{gw},0}}$}
  allowed for signals with fixed \mbox{$f_{\mathrm{gw},0} = 2000$\,Hz}
  and the other model parameters taking values in the intervals
  \mbox{$\tau \in [1000, 9640]$\,s}
  and \mbox{$n \in [2.5, 7]$}.
 }
\end{figure}

We first obtain a collection of SFTs by dividing the full observation time $T_\mathrm{obs}$ in $N$ segments of length $T_\mathrm{coh}$.
The maximum length of $T_\mathrm{coh}$ is calculated by imposing the 1/4-cycle criterion introduced in \cite{grav1}:
This leads to a requirement $2|df/dt|\leq T_\mathrm{coh}^{-2}$,
\updated{ensuring that the maximum modulation corresponds to only half a bin
at the search frequency resolution \mbox{$\delta f=1/T_\mathrm{obs}$}.}
From Eq.~\eqref{doppler} the spin-down modulation is given by two effects,
the spin-down of the source and the Doppler modulation resulting from the Earth's motion.
The constraint imposed by the spin-down of the source is:
\begin{equation}
 \label{tcohe}
 T_\mathrm{coh}\leq\frac{\sqrt{(n-1) \tau }}{\sqrt{2 f_{\mathrm{gw},0}}} \,.
\end{equation}

The range of maximum allowed $T_\mathrm{coh}$ for the parameter space covered in \cite{GW170817remnant}
is on the order of seconds,
as shown in Fig.~\ref{fig:tcoh}.
On the other hand, the constraint imposed by Doppler modulation is on the order of hours,
as discussed in \cite{Hough}.
Therefore we will consider only the spin-down of the source as the dominant threshold for $T_\mathrm{coh}$.

\begin{figure*}
 \includegraphics[width=\columnwidth]{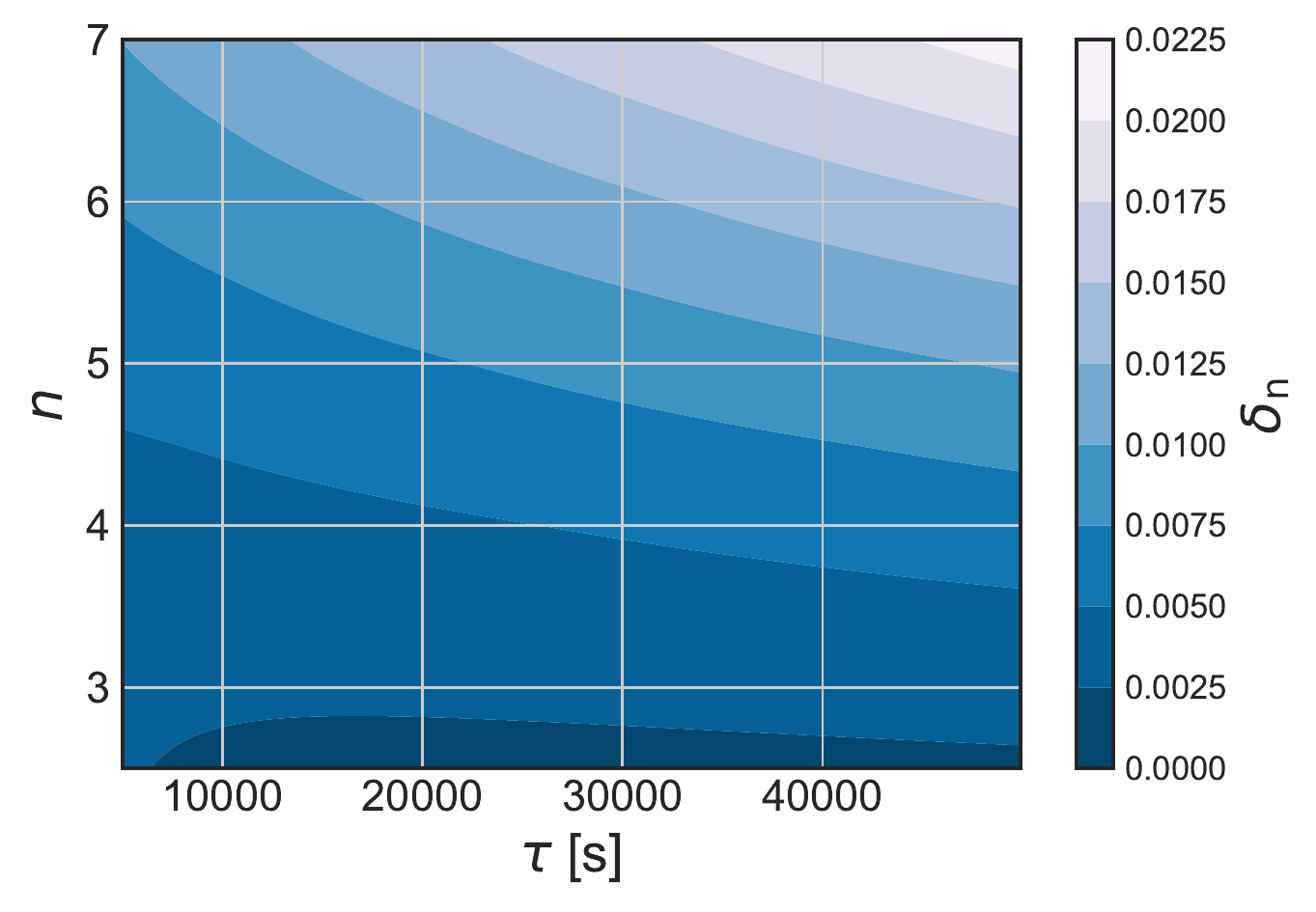}
 \includegraphics[width=\columnwidth]{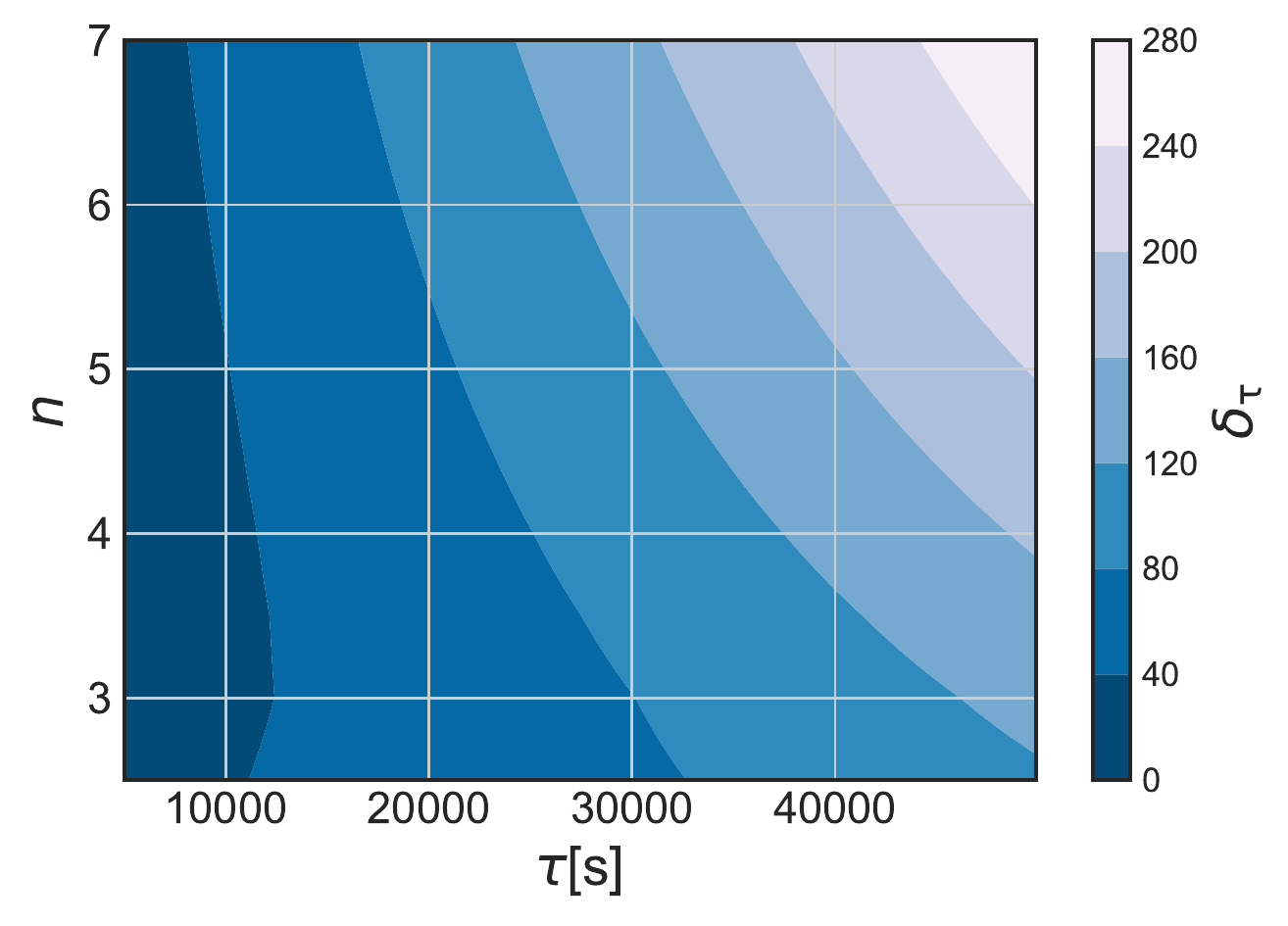}
 \vspace{-\baselineskip}
 \caption{
  \label{Grid_contours}
  Non-uniform search grid setup:
  step sizes $\delta n$ (left panel)
  and $\delta \tau$ (right panel)
  in the braking index and spin-down timescale
  obtained by setting
  \mbox{$T_\mathrm{obs}=86400$\,s},
  \mbox{$T_\mathrm{coh}=1$\,s},
  fixed \mbox{$f_{\mathrm{gw},0} = 2000$\,Hz}
  (corresponding to the maximum of the search range),
  and as a function of \mbox{$\tau \in [1000, 9640]$\,s}
  and \mbox{$n \in [2.5, 7] $}.
  In practice, while we will follow Eq.~\eqref{dtau} to select $\delta \tau$ at each step
  (finer grid at smaller $\tau$),
  we will always select the minimum value of $\delta n$ (finest grid) in a given parameter range.
 }
\end{figure*}

\subsection{The peak-gram}

The Hough transform requires a digitized spectrum as its input,
with time-frequency bins categorized in two classes.
The ATrHough generates this by setting a threshold $\rho_\mathrm{th}$ on the normalized power spectrum $\rho_i$ to conduct the bin selection:
\begin{equation}
 \label{normpower}
 \updated{\rho_{i,k}} \approx \frac{2|\tilde{x}_i[f_k]|^2}{T_\mathrm{SFT} S_\mathrm{n}[f_k]} \,,
\end{equation}
where $[.]$ indicates a discrete series
and the index $i$ corresponds to the $i^\mathrm{th}$ time step.
That is, $\tilde{x}_i[f_k]$ is the value obtained from the $i^\mathrm{th}$ SFT on the $k^\mathrm{th}$ frequency bin.
Furthermore, $S_\mathrm{n}$ is the single-sided Power Spectral Density (PSD) of the noise in the same bin.
\updated{In the following, we drop the explicit $k$ index,
as we are only interested in the frequency bins following the signal track.}
If $\rho_i \geq \rho_\mathrm{th}$,
then a value of $1$ is assigned to that bin, and a $0$ otherwise.
The result of this process is known as the peak-gram.

\subsection{Resolution in \texorpdfstring{$\tau$}{tau} and \texorpdfstring{$n$}{n} space}

The Hough transform is applied to find the statistical significance of each template in a bank over parameter space.
A template is defined by the intrinsic parameters of the signal, \mbox{$\vec{\xi}=(f_{\mathrm{gw},0},n,\tau,T_{0})$}.
To conduct a wide-parameter space search,
we create a grid that ensures contiguous templates to deviate from each other by at most one frequency bin over a duration $T_\mathrm{obs}$;
this ensures the computation of at least all independent templates (by the 1/4-cycle criterion) between \mbox{$t=0$\,s} and \mbox{$t=T_\mathrm{obs}$}.
The grid is constructed with the following step sizes:
\begin{subequations}
\begin{align}
\delta n &= \frac{\partial n}{\partial \fgw(t)}\Big|_{t=T_\mathrm{obs}} \delta f \,, \\
\delta \tau &= \frac{\partial \tau}{\partial \fgw(t)}\Big|_{t=T_\mathrm{obs}} \delta f \,,
\end{align}
\end{subequations}
where $\delta f =1/T_\mathrm{coh}$.
Hence,
\begin{subequations}
\begin{align}
 \label{dbrakindex}
 \delta n    &= \frac{(n-1)^2 \left(\frac{T_\mathrm{obs}}{\tau }+1\right)^{-\frac{1}{1-n}}}{f_{\mathrm{gw},0} T_\mathrm{coh} \log \left(\frac{T_\mathrm{obs}}{\tau }+1\right)} \,, \\
 \label{dtau}
 \delta \tau &= \frac{(n-1) \tau  (\tau +T_\mathrm{obs}) \left(\frac{T_\mathrm{obs}}{\tau }+1\right)^{-\frac{1}{1-n}}}{f_{\mathrm{gw},0} T_\mathrm{coh} T_\mathrm{obs}} \,.
\end{align}
\end{subequations}

\begin{figure}
 \includegraphics[width=\columnwidth]{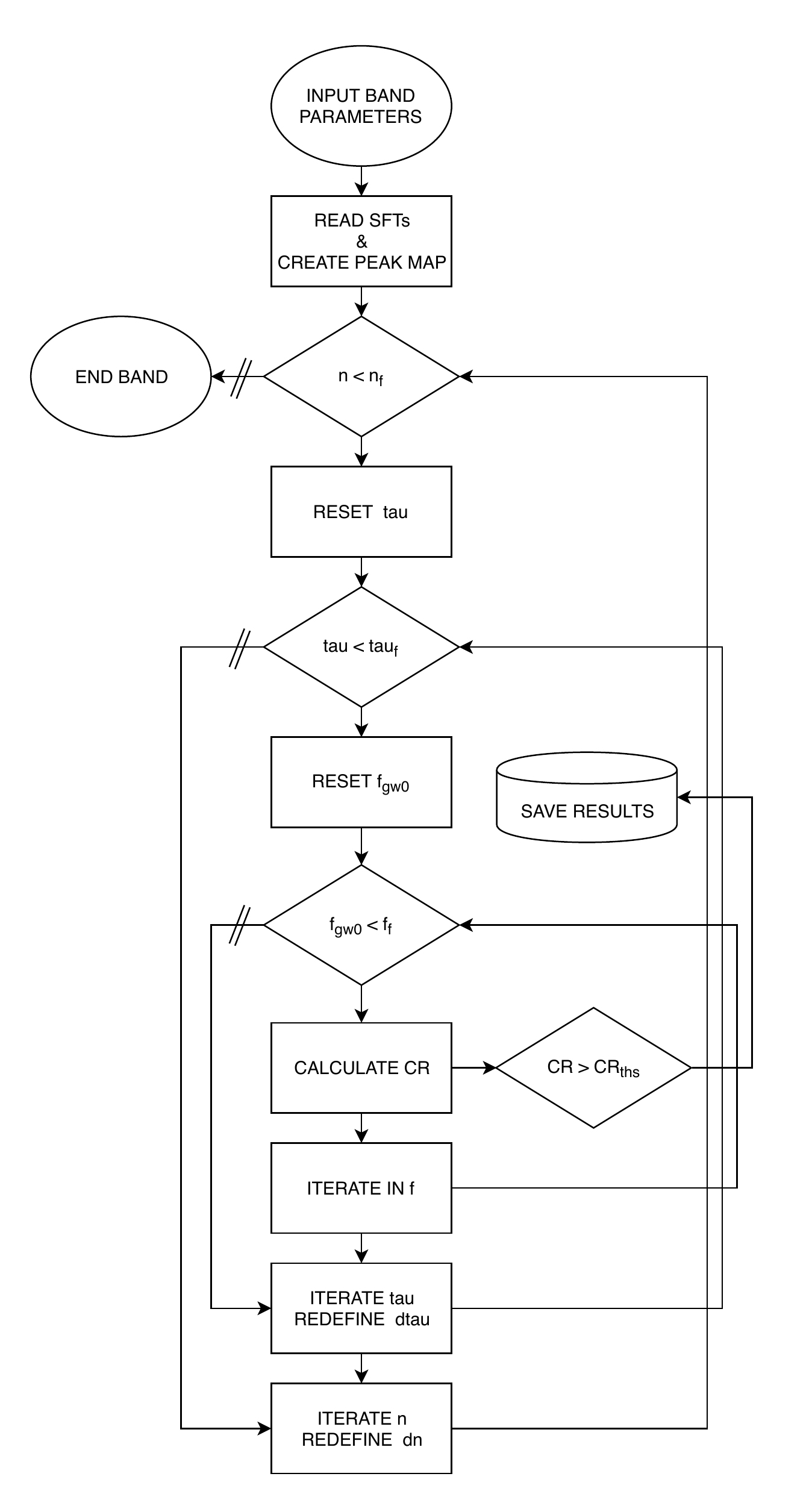}
 \vspace{-\baselineskip}
 \caption{
  \label{Pipeline_diagram}
  A diagram of the ATrHough work flow inside a single search band.
  Arrows indicate the stream direction,
  squares correspond to input/output calculations
  and diamonds to if-statements with double lines indicating a `false' outcome.
  The entire pipeline includes multiple calls to calculate all bands inside the parameter-space domain.
 }
\end{figure}

The two grid step sizes are inversely proportional to $f_{\mathrm{gw},0}$.
Fig.~\ref{Grid_contours} represents
the obtained $\delta \tau$ and $\delta n$
for a fixed $T_\mathrm{coh}$, $T_\mathrm{obs}$ and $f_{\mathrm{gw},0}$
inside the $\tau$, $n$ ranges.

\begin{figure}
 \includegraphics[width=\columnwidth]{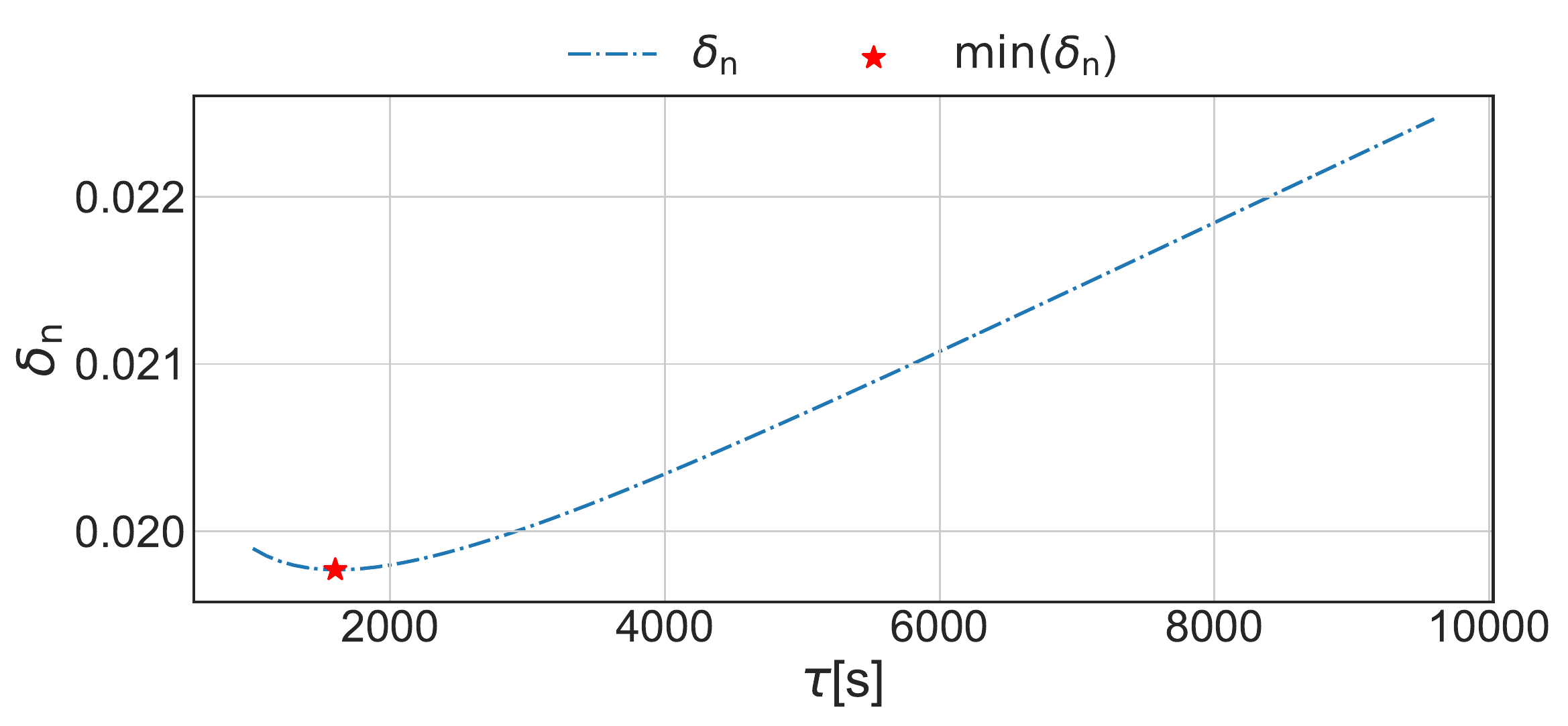}
 \vspace{-\baselineskip}
 \caption{
  \label{Grid_dn}
  Example of the grid step size $\delta n$
  as a function of \mbox{$\tau \in [1000, 9640]$\,s},
  obtained by setting
  \mbox{$n=5$},
  \mbox{$T_\mathrm{obs}=86400$\,s},
  \mbox{$T_\mathrm{coh}=1$\,s},
  and for a frequency range with maximum \mbox{$f_{\mathrm{gw},0} = 550$\,Hz}.
  The red star corresponds to \mbox{$\delta n_\mathrm{min}$},
  which in the practical search implementation we pick as a fixed value for this parameter range.
 }
\end{figure}

\begin{figure}
 \includegraphics[width=\columnwidth]{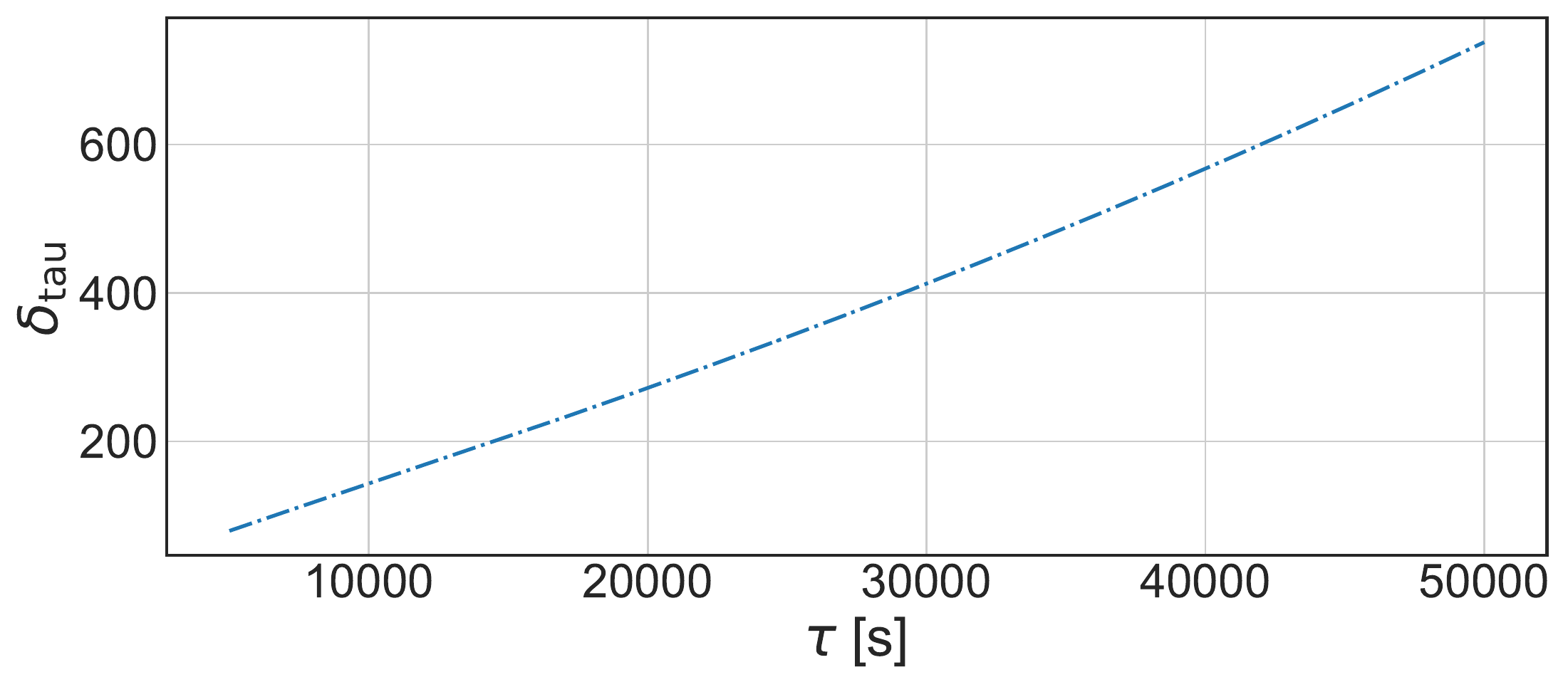}
 \vspace{-\baselineskip}
 \caption{
  \label{Grid_dtau}
  Example of the grid step size $\delta n$
  as a function of \mbox{$\tau \in [1000, 9640]$\,s},
  obtained by setting
  \mbox{$n=5$},
  \mbox{$T_\mathrm{obs}=86400$\,s},
  \mbox{$T_\mathrm{coh}=1$\,s},
  and for a frequency range with maximum \mbox{$f_{\mathrm{gw},0} = 550$\,Hz}.
 }
\end{figure}

The practical implementation of the grid is defined by a nested loop;
a pipeline diagram can be seen in Fig.~\ref{Pipeline_diagram}.
First, we select the minimum value of $\delta n$ over the $\tau$ range as shown in Fig.~\ref{Grid_dn},
given a set of $(T_\mathrm{obs}, T_\mathrm{coh}, n)$ and the maximum $f_{\mathrm{gw},0}$;
then we calculate $\delta \tau$ as in Fig.~\ref{Grid_dtau}.
We will recalculate $\delta n$ and $\delta \tau$ on each iteration of the $n$ and $\tau$ loops respectively.

In order to reduce the number of templates or grid points required by the search,
we need to split the $\tau$ and $f_{\mathrm{gw},0}$ ranges of the whole search space into smaller subdomains.
To do so, we will typically create bands for $\tau$ smaller than $10\%$ of $T_\mathrm{obs}$
and frequency bands between 50 and 100\,Hz in width.
Each sub-domain will be analyzed independently, making the computational load smaller.
It is possible to make the domains larger,
but the necessary refinement of the grid in certain areas will make the search less computationally efficient overall.

\begin{figure}
 \includegraphics[width=0.95\columnwidth]{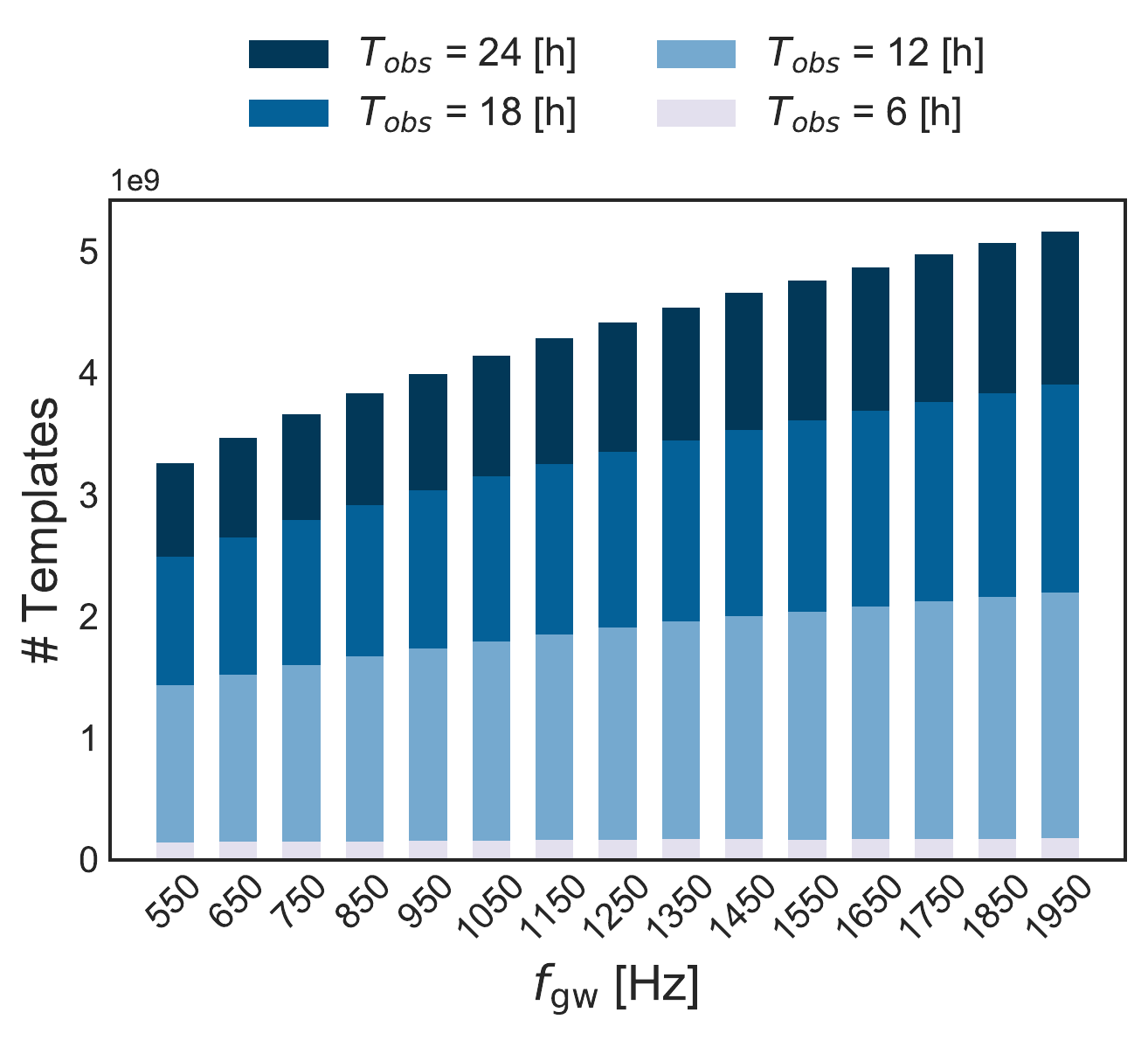}
 \includegraphics[width=0.95\columnwidth]{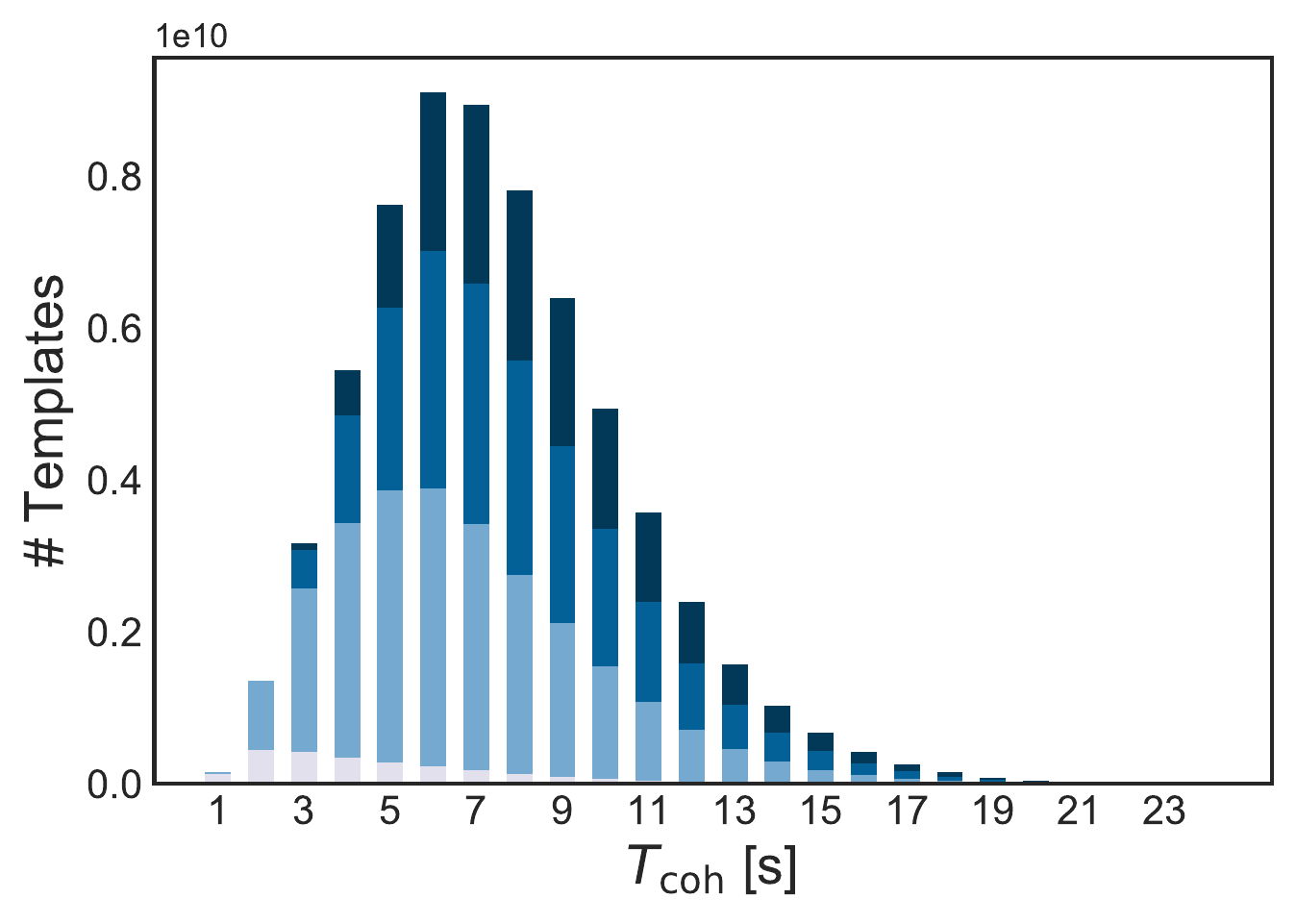}
 \includegraphics[width=0.95\columnwidth]{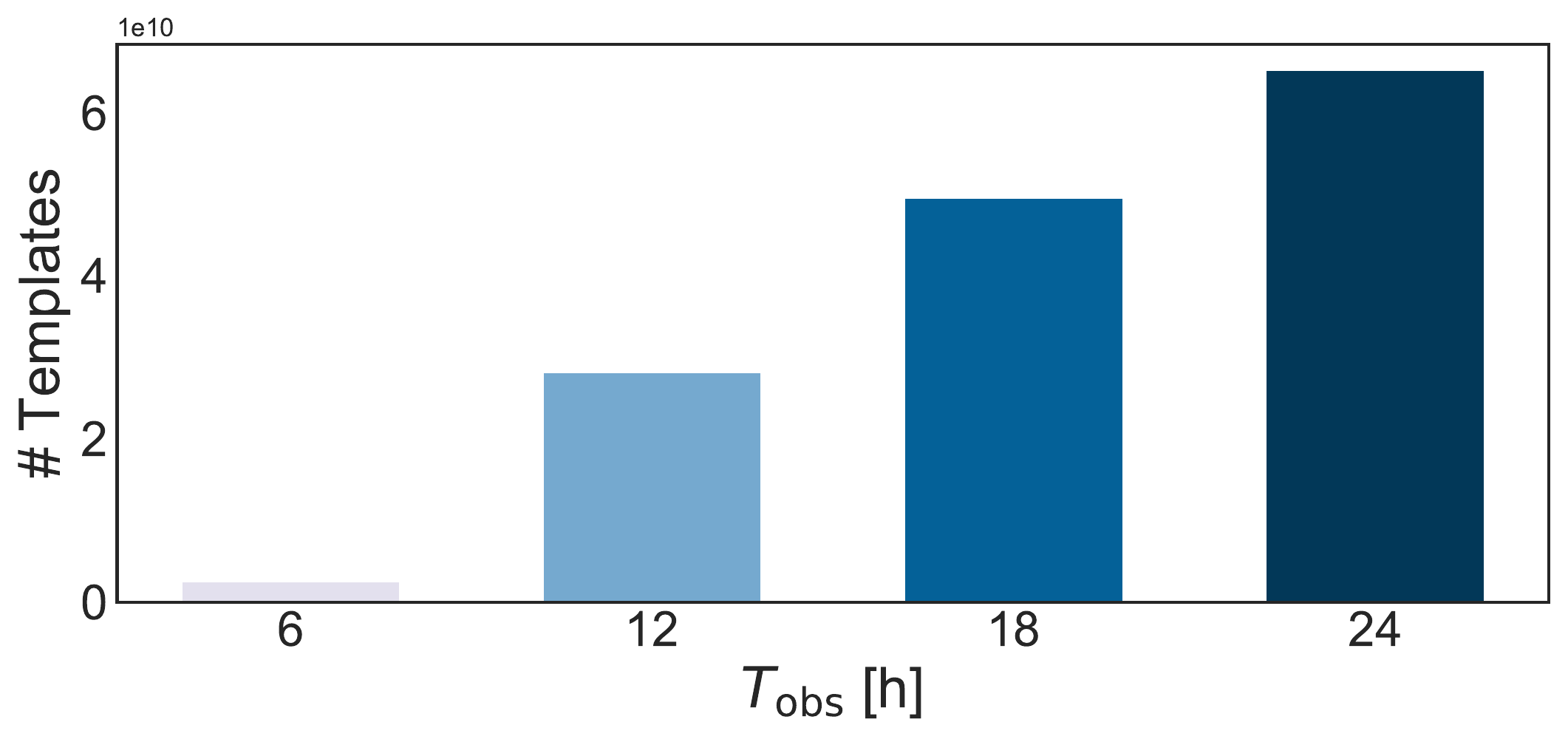}
 \vspace{-\baselineskip}
 \caption{
  \label{Ntemplates}
  The number of templates required for searches with four different $T_\mathrm{obs}$.
  The total parameter-space covered is
  \mbox{$n \in [2.5,7]$},
  \mbox{$f_{\mathrm{gw},0} \in [500,2000]$\,Hz},
  \mbox{$\tau \in [10^3,10^5]$\,s}
  and is evaluated in independently-processed subdomains,
  each corresponding to a $\tau$ band of $100$\,s
  and a 100\,Hz wide frequency band.
  In this figure, all panels show counts of templates after combining the $\tau$ bands.
  The top panel shows the number of templates for each frequency band
  when using the optimal $T_\mathrm{coh}$ for each $T_\mathrm{obs}$:
  it increases with $f_{\mathrm{gw},0}$ for each $T_\mathrm{obs}$,
  and longer $T_\mathrm{obs}$ require more templates at each frequency.
  The middle panel shows the total number of templates (summed over all frequency bands),
  for each $T_\mathrm{obs}$,
  as a function of $T_\mathrm{coh}$.
  The lower panel shows the total number of templates
  when again using the optimal $T_\mathrm{coh}$ for each $T_\mathrm{obs}$.
 }
\end{figure}

Fig.~\ref{Ntemplates} shows the distribution and number of templates used for different $T_\mathrm{obs}$
given a search that covers an analogous parameter space as \cite{GW170817remnant}.
Here templates are calculated with the maximum integer coherence length allowed,
and the minimum $T_\mathrm{coh}$ considered for this figure and the search is 1\,s.

\section{STATISTICAL PROPERTIES}

\subsection{The coherent statistic}

For the following section we make the assumption of stationary Gaussian noise with 
zero mean in order to characterize the output of the detectors,
for which the normalized power $2\rho_{i}$ in the presence of a signal $h$ 
follows a non-central $\chi^2$ distribution with 2 degrees of freedom
and a non-centrality parameter
\begin{equation}
 \lambda_{i}=\frac{4|\widetilde{h_i}[f_k]|^2}{T_\mathrm{SFT}S_n[f_k]} \,,
\end{equation}
where $|\widetilde{h_i}[f_k]|$ is the Fourier transform of the signal
\updated{and, as before in Eq.~\eqref{normpower} for the normalized power $\rho_i$,
for $\lambda_{i}$ we suppress the $k$ dependence.}
Then the probability distribution for $\rho_i$ is:
\begin{equation}
p(\rho_i|\lambda_i) = 2 \chi^2(2 \rho_i|2,\lambda_i) = \exp(-\rho_i-\frac{\lambda_i}{2})I_0(\sqrt{2\lambda_i\rho_i}) \,,
\end{equation}
where $I_0$ is the zero-order modified Bessel function of the first kind.

The mean and variance for this distribution are respectively:
\begin{subequations}
\begin{eqnarray}
\textbf{E}[\rho_i] &=& 1+\frac{\lambda_i}{2} \,,\\
\sigma^2 [\rho_i] &=& 1+\lambda_i \,.
\end{eqnarray}
\end{subequations}

The false alarm and false dismissal probabilities for a frequency bin to be above the power spectrum threshold are:
\begin{subequations}
\begin{eqnarray}
\alpha(\rho_\mathrm{th})&=&\int_{\rho_\mathrm{th}}^{\infty}p(\rho|0)d\rho = \exp(-\rho_\mathrm{th}),\\
\beta_\mathrm{i}(\rho_\mathrm{th})&=&\int_{0}^{\rho_\mathrm{th}}p(\rho|\lambda_i)d\rho = 1 - \eta_i(\rho_\mathrm{th}|\lambda_i).
\end{eqnarray}
\end{subequations}
The probability $\eta_i$ that a given frequency bin is selected is,
in the small-signal approximation:
\begin{equation}
\eta_i(\rho_\mathrm{th}|\lambda_i)=\int_{\rho_\mathrm{th}}^{\infty}p(\rho|\lambda)d\rho=\alpha \Big(1+\frac{\rho_\mathrm{th}}{2}\lambda_i+O(\lambda_i^2)\Big).
\end{equation}

\subsection{The incoherent number-count statistic}

If a signal is present, the non-centrality parameter $\lambda_i$ will change for different SFTs. 
As pointed out previously, this can happen both
because the noise may not be stationary
and because the amplitude modulation of the signal changes over time.
In other words, the observed signal power $|h|^2$ changes due to the non-uniform 
antenna pattern of the detector and due to the intrinsic spindown. 
Therefore, the detection probability \updated{$\eta_i$} changes across SFTs.
This is taken into account by \updated{adapting} the non-demodulated weighted Hough approach mentioned before and covered in \cite{WHough};
it is a similar strategy to the one applied in the StackSlide \citep{stackslide} and PowerFlux \citep{power1,power2} algorithms.
The starting point is to generalize the integer number-count statistic,
which we would obtain directly from the peak-map,
to a non-integer weighted statistic
\begin{equation}
 \nu = \sum_{i=1}^{N} w_i \, \nu_i \,,
 \label{nsum}
\end{equation}
where $N$ is the number of SFTs,
$\nu_i$ is the value assigned to the bin selected from the peak-gram 
in the $i^\mathrm{th}$ time step for the current template,
and \updated{$w_i$ are a constant set of weights given for each template with \mbox{$w_I \propto 1/N$}. 
It is important to notice that in order to maximize the sensitivity of the search the 
selection of weights is not arbitrary;
we will derive the optimal choice in Sec.~\ref{sec:calibration}.
For now, we define the normalization terms}
\begin{subequations}
 \label{wsums}
 \begin{eqnarray}
  A       = \sum_{i=1}^{N} w_i \,, \label{Asum} \\
  ||w||^2 = \sum_{i=1}^{N} w_i^2 \,, \label{w2sum}
  \end{eqnarray}
\end{subequations}

\updated{This step in the search (computing $\nu$)} is known as the incoherent sum;
the templates in a search are then ranked based on their number count $\nu$.
\updated{Applying the linearity of the expectation value, the mean and variance for the incoherent 
step in the absence of a signal are:}
\begin{subequations}
\label{mean-variance}
\begin{align}
\langle \nu \rangle   &= A \, \alpha \,, \label{nuavg} \\
\sigma_\mathrm{\nu}^2 &= \langle \nu^2 \rangle-\langle \nu \rangle^2 =||w||^2\,\alpha\,(1-\alpha)\,. \label{sigmamma}
\end{align}
\end{subequations}

\updated{As shown in \cite{WHough} and applied in multiple \cw searches like \cite{Astone:2014esa},
when optimal weights are chosen we can,
for a sufficient number of SFTs,
evaluate the significance of an observation by
approximating the number count distribution by a Gaussian with the right mean and variance:}
\updated{
 \begin{equation}
 p(\nu|\rho_\mathrm{th},\lambda) = \frac{1}{\sqrt{2\pi\sigma^2}}e^{-(\nu-A\alpha)^2/2\sigma^2} \,.
 \label{gaussianapprox}
\end{equation}
}
\updated{This becomes a very good approximation for $N>1000$,
and e.g. the typical number of SFTs searched in \cite{GW170817remnant}
is indeed above that number.
\newupdated{We provide some empirical tests of this approximation in appendix \ref{sec:gaussiankl}.}
}

Thus one can derive the number count threshold $\nu_{th}$ based on the incoherent false-alarm rate as
\begin{equation}
 \label{phithreshold1}
 \alpha_\mathrm{I} = \int_{\nu_\mathrm{th}}^{\infty}p(\nu|\rho_\mathrm{th},0)d\nu
                   = \frac{1}{2}\erfc\Big(\frac{\nu_\mathrm{th}-\langle \nu \rangle}{\sqrt{2}\sigma_\mathrm{\nu}} \Big) \,.
\end{equation}
For a given set of weights and peak selection threshold,
this equation decides what number count threshold must be used to obtain a desired $\alpha_\mathrm{I}$.
We can solve this as
\begin{equation}
 \label{phithreshold}
 \nu_\mathrm{th} = A \alpha + \sqrt{2||w||^2\alpha(1-\alpha)}\,\erfc^{-1}(2\alpha_\mathrm{I}) \,.
\end{equation}

\updated{The false-dismissal rate requires the computation of the mean and variance,
which in the presence of a small signal are:}
\updated{\begin{subequations}\begin{eqnarray}
\langle \eta \rangle &=& \sum_{i=1}^Nw_i\eta_i \sim A\alpha+\frac{\alpha \rho_\mathrm{th}}{2}\sum_{i=1}^N w_i\lambda_i,\\
\sigma_\mathrm{\eta}^2&=&\sum_{i=1}^Nw_i^2\eta_i(1-\eta_i).
\end{eqnarray}\end{subequations}}
\updated{If the small-signal approximation is applied,
$\sigma_\mathrm{\eta}^2$ can be expanded to first order in $\lambda_i$:}
\updated{\begin{equation}\label{nalpha}
\sigma_{\eta}^2=||w||^2\alpha(1-\alpha)\Big(1+\frac{\rho_\mathrm{th}}{2||w||^2}\frac{1-2\alpha}{1-\alpha}\sum_{i=1}^Nw_i^2\lambda_i\Big).
\end{equation}}

We again approximate the number count distribution \updated{$p(\eta|h)$} by a Gaussian distribution
with the above mean and variance, yielding the false-dismissal rate as follows:
\updated{\begin{equation}
\beta_\mathrm{I}\approx\int_{-\infty}^{\nu_\mathrm{th}}p(\eta|h)dn = \frac{1}{2}\erfc\Big(\frac{\langle \eta \rangle-\nu_\mathrm{th}}{\sqrt{2}\sigma_\mathrm{\eta}}\Big).
\end{equation}}

\subsection{Setting up the threshold}

Considering the statistical significance in a template as
\mbox{$\,s := 1 - \alpha_\mathrm{I} - \beta_\mathrm{I}$}
and using the properties of the complementary error function,
we can introduce a quantity
\begin{equation}
 S=\erfc^{-1}(2\alpha_\mathrm{I})+\erfc^{-1}(2\beta_\mathrm{I}) \,.
\end{equation}
This equation can be shown to reduce to $s$ when \mbox{$S=0$},
and as it grows monotonically we can take it as a measure of the statistical significance of the search.
By expanding to the first order in $\lambda_i$,
we derive the following expression:
\begin{equation}
 \begin{split}
  S =&\; \sqrt{\frac{\alpha\rho_\mathrm{th}^2}{8(1-\alpha)}}\frac{\sum_{i=1}^Nw_i\lambda_i}{||w||} \\
     &+\frac{\rho_\mathrm{th}}{4}\frac{1-2\alpha}{1-\alpha}\frac{\sum_{i=1}^Nw_i\lambda_i}{||w||^2}\,\erfc^{-1}(2\alpha) \,.
 \end{split}
\end{equation}

\updated{Imposing again optimal weights which are proportional to $1/N$,
for large values of $N$ the first term on the right-hand side of this equation is proportional to $\sqrt{N}$,
while the second term does not grow with $N$.}
Thus the first term dominates, yielding
\begin{equation}
 \label{sensbefore}
 S \sim \sqrt{\frac{\alpha\rho_\mathrm{th}^2}{8(1-\alpha)}}\frac{\sum_{i=1}^Nw_i\lambda_i}{||w||} \,.
\end{equation}

The peak selection threshold is chosen to minimize $\beta_\mathrm{I}$,
or equivalently maximize $S$ for fixed $\alpha_\mathrm{I}$:
\begin{equation}
 \frac{d}{d\rho_\mathrm{th}}\sqrt{\frac{\alpha\rho_\mathrm{th}^2}{8(1-\alpha)}}=0 \,.
\end{equation}
\updated{As derived in \cite{Hough}, this threshold is independent of the choice of weights;
and the solution of the previous equation is \mbox{$\rho_\mathrm{th} = 1.6$}
which leads to \mbox{$\alpha = e^{-\rho_\mathrm{th}}= 0.2$}.
Different thresholds can be imposed,
yielding different \mbox{$\alpha$},
but they would not maximize the
statistical significance of the template.}

\subsection{Calibration of the weights}
\label{sec:calibration}

To define an appropriate set of weights,
we start by considering the modulus square of the signal's Fourier transform on the $i^\mathrm{th}$ SFT,
\updated{depending on the antenna patterns $F_{+,\times}$ from Eq.~\eqref{eq:FplusFcross} and amplitudes $A_{+,\times}$ from Eq.~\eqref{eq:AplusAcross}}:
\updated{\begin{equation}
 \hspace{-1cm}
 |\widetilde{h_i}[f_k]|^2=\frac{A_{+,i}^2F_{+,i}^2+ A_{+,i}^2F_{\times,i}^2}{4} \frac{\sin^2[\pi(f_{\mathrm{gw},i}-f_k)T_\mathrm{coh}]}{\pi(f_{\mathrm{gw},i}-f_k)}\,.
\end{equation}}
\updated{From here on, the subindex $i$ runs over segments and in the case of a fuction
it imposes a time average of length $T_\mathrm{coh}$,
e.g. for the time-evolving \gw frequency from Eq.~\eqref{model}:
\mbox{$f_{\mathrm{gw},i} = \int_{T_i-T_\mathrm{coh}/2}^{T_i+T_\mathrm{coh}/2} \fgw(t)dt/T_\mathrm{coh}$}.
The subindex $k$ corresponds to the $k^\mathrm{th}$ frequency bin, selected so that
\mbox{$f_{\mathrm{gw},i} \in (f_k - \delta f/2, f_k+ \delta f/2)$}.}
The average over that interval is
\begin{equation}
\int^{\frac{1}{2}}_{-\frac{1}{2}}\frac{\sin^2[\pi x]}{(\pi x)^2}=0.7737 \,.
\end{equation}

Now we can average over the \updated{\ns['s]} orientation $\cos\iota$ and the polarization angle $\psi$ appearing in the antenna patterns
and find the following relationships:
\begin{subequations}
\begin{eqnarray}
\left<(F_{+,i})^2\right>_{\iota,\Phi}     = \left<(F_{\times,i})^2\right>_{\iota,\psi} &=& \frac{a_i^2+b_i^2}{2} \,, \\
\left<(A_{+,i})^2+(A_{\times,i})^2\right>_{\iota,\psi} &\sim & \frac{4 h_{0,0}^2}{5}\Big(\frac{f_{\mathrm{gw},i}}{f_{\mathrm{gw},0}}\Big)^{2m} \,,
\end{eqnarray}
\end{subequations}
where \mbox{$h_{0,0}=h_0(t=t_0)$} is the initial amplitude at $t_0$.

Combining all these results:
\begin{equation}
 \left<\lambda_i\right>_{\iota,\psi} = 0.7737 \frac{2 h_{0,0}^2 T_\mathrm{coh}(a_i^2+b_i^2)}{5S_{\mathrm{n},i}}\Big(\frac{f_{\mathrm{gw},i}}{f_{\mathrm{gw},0}}\Big)^{2m} \,,
\end{equation}
and substituting this into Eq.~\eqref{sensbefore},
the sensitivity is
\begin{equation}
 \label{sensitivity}
 S = \sqrt{\frac{\alpha\rho_\mathrm{th}^2}{8(1-\alpha)}}\frac{2h_{0,0}^2T_\mathrm{coh}}{5||w||}
     \sum_{i=1}^N w_i\frac{(a_i^2+b_i^2)}{S_{\mathrm{n},i}}\Big(\frac{f_{\mathrm{gw},i}}{f_{\mathrm{gw},0}}\Big)^{2m} \,.
\end{equation}

\updated{From this, we see that the sensitivity is related to 
the detector response and the amplitude modulation of the signal,
which we can summarize in a quantity}
\begin{equation}
X_i := \frac{(a_i^2+b_i^2)}{S_{\mathrm{n},i}}\Big(\frac{f_{\mathrm{gw},i}}{f_{\mathrm{gw},0}}\Big)^{2m} \,.
\end{equation}

\updated{Calculating the maximum of the inner product \mbox{$\textbf{w} \cdot \textbf{X}$} 
shows that the weights guarantee the best sensitivity for a given template
if the two vectors are proportional to each other, i.e. $w_i \propto X_i$.
At the same time, we see that any overall rescaling of the weights \mbox{($\hat{w}_i = k w_i$)}
has no impact on $S$, as for any constant $k$ the value of the detectable dimensionless 
strain amplitude $h_\mathrm{0,0}$ at \mbox{$t=0$\,s} remains unchanged.}

\begin{figure}
 \includegraphics[width=\columnwidth]{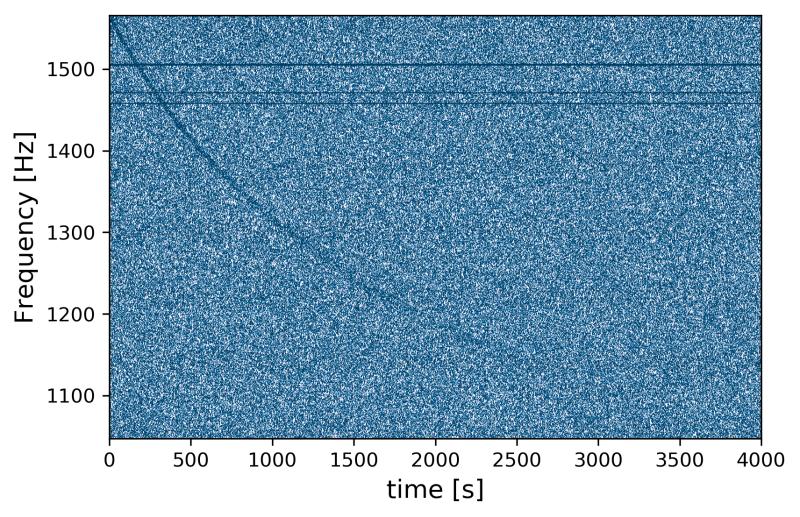}
 \includegraphics[width=\columnwidth]{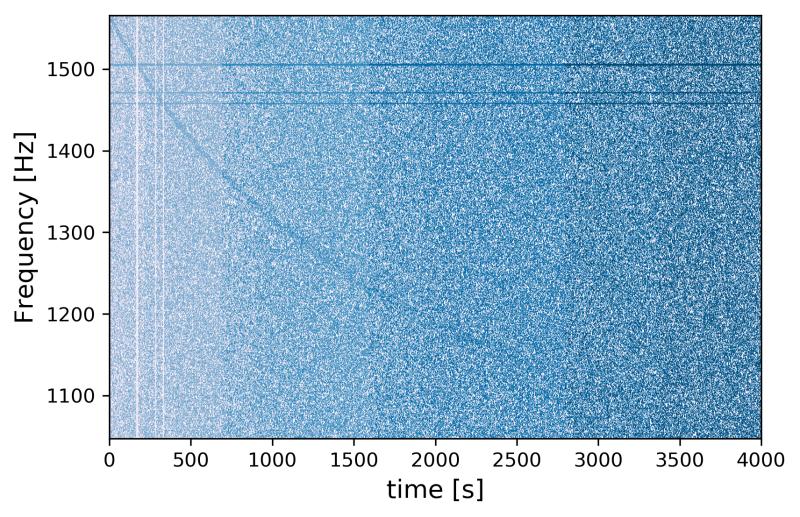}
 \includegraphics[width=\columnwidth]{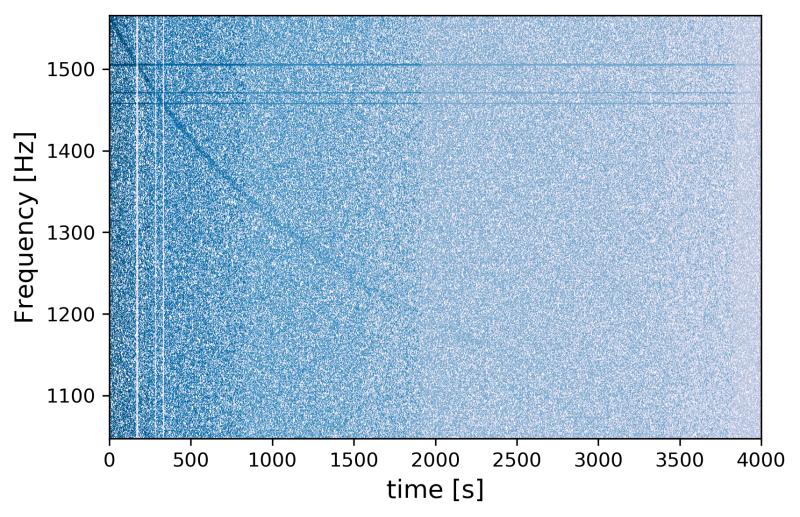}
 \includegraphics[width=0.7\columnwidth]{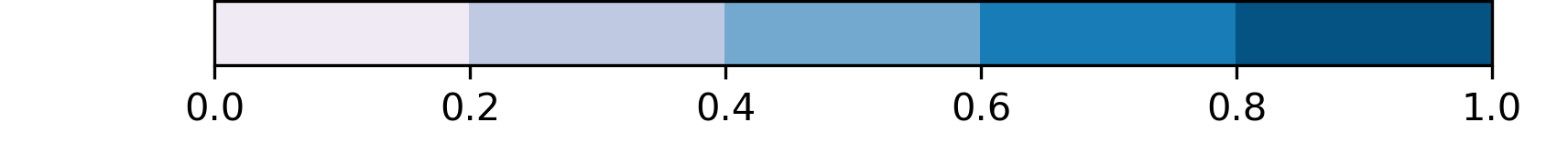} 
 \vspace{-\baselineskip}
 \caption{
 \label{injection}
 \updated{
  Peak-maps for an example simulated signal at \mbox{$0.1$\,Mpc} using
  \mbox{$T_\mathrm{coh}=1$\,s} in the actual aLIGO data after GW170817
  with NS parameters \mbox{$f_{gw,i} = 1565.8$\,Hz}, \mbox{$\tau = 1000$\,s}, \mbox{$n = 5$},
  \mbox{$I_\mathrm{zz}=4.34 \times 10^{38}$\,kg\,m$^2$} and \mbox{$\cos\iota=1$}.
  The top panel does not use weights,
  the middle panel uses weights as in \cite{WHough} which do not include the source amplitude modulation,
  and the bottom panel corresponds to the new weights derived in Sec.~\ref{sec:calibration}.
  The color scale is $w_i\nu_i$, normalized to be comparable between panels.
  In all cases, the signal track disappears below the noise floor at around 2000\,s,
  as expected from the injection parameters and the detector noise curve.
  In the first two panels we see that noise contributions from the later part
  of the observing window will decrease significance with time.
  However in the bottom panel, these are weighted down,
  increasing the significance of the recovered track.
  }
 }
\end{figure}

\updated{In summary,
as also illustrated for an example simulated signal in Fig.~\ref{injection},
the use of appropriate weights ensures our search properly accounts for both
the source's amplitude decay
and the effects of the detector response changing with time.
This gives us the ability to compare templates across the search parameter space,
comparing very fast frequency decays with slower ones.}

If the value $\rho_\mathrm{th}$ = 1.6 is substituted in Eq.~\eqref{sensitivity},
the minimum theoretical value \updated{of $h_0$ that the search can recover} is:
\begin{equation}
 \label{sensitivityfunct}
 h_{0,0}=3.38 \sqrt{\frac{S^{1/2}}{T_\mathrm{coh}}} \Big(\frac{||\textbf{w}||}{\textbf{w} \cdot \textbf{X}}\Big)^{(1/2)}.
\end{equation}

\subsection{Critical Ratio \texorpdfstring{$\Psi$}{Psi}}

The critical ratio $\Psi$ is a
new statistic
that quantifies the significance of a given template.
Based on the weighted number count and quantities from Eqs.~\eqref{nsum}--\eqref{mean-variance},
we define
\begin{eqnarray}
 \Psi &=& \frac{\nu-\langle \nu \rangle}{\sigma_\nu^2} \nonumber\\
      &=& \frac{\sum_{i=1}^{N}(w_i \nu_i)-\sum_{i=1}^{N}(w_i\alpha)}{\sqrt{\sum_{i=1}^{N}(w_i)^2\alpha(1-\alpha)}}. \,
\end{eqnarray}

As mentioned before, any normalization of the weights will not change the sensitivity of the search.
It will also leave the significance or critical ratio in each template unchanged.
Considering the previous equation as the single-detector case,
the multi-detector critical ratio is defined as
\begin{equation}
 \label{criteicalm1}
 \Psi_\mathrm{M} = \frac{\sum_{k=1}^{N_\mathrm{M}}(\sum_{i=1}^{N_k}(w_{i;k} \nu_{i;k})-\sum_{i=1}^{N_k}(w_{i;k}\alpha))}
                        {\sqrt{ \sum_{k=1}^{N_\mathrm{M}}\sum_{i=1}^{N_k} (w_{i;k})^2\alpha(1-\alpha)}} \,,
\end{equation}
where $N_\mathrm{M}$ is the number of detectors
and $N_k$ is the number of SFTs in detector $k$,
while $w_{i;k}$ and $\nu_{i;k}$ are the weights and number count assigned to the $i^{th}$ SFT for that detector and a given template.
We can also rewrite this as
\begin{equation}
 \Psi_\mathrm{M} = \frac{\sum_{k=1}^{N_\mathrm{M}} \Psi_k \sqrt{\sum_{i=1}^{N_k} (w_{i;k})^2 }}
                        {\sqrt{\sum_{k=1}^{N_\mathrm{M}} \sum_{i=1}^{N_k} (w_{i;k})^2}},
\end{equation}
where $\Psi_k$ is the critical ratio for each single detector $k$.

In a multi-detector search,
the duty factors
(fraction of time a detector is recording usable data)
and noise floors may differ between detectors.
To quantify the contribution of each detector to the multi-detector critical ratio,
the relative contribution ratio is defined as
\begin{equation}
 r_{j}=\sqrt{\frac{\sum_{i=1}^{\nu_j} (w_{i;j})^2}{\sum_{k=1}^{N_\mathrm{M}}\sum_{i=1}^{\nu_k} (w_{i;j})^2}} \,.
\end{equation}

Using the previous equations,
the critical ratio for a multi-detector search takes a very simple form:
\begin{equation}\label{multiifo}
 \Psi_\mathrm{M} = \sum_{k=1}^{N_\mathrm{M}} \Psi_k r_{k} \,.
\end{equation}

\section{Vetoes on Critical ratio and Time Consistency}

Candidates that appear significant by their critical ratio can be due to astrophysical sources,
but also due to non-Gaussian noise artifacts in the data.
To make the search robust against such artifacts,
we introduce vetoes that test for each candidate
(i) its consistency between detectors
and (ii) the consistency of its transient behavior with the target astrophysical model.

\subsection{The Critical ratio \texorpdfstring{$\Psi$}{Psi}-veto}

\begin{figure}
 \includegraphics[width=\columnwidth]{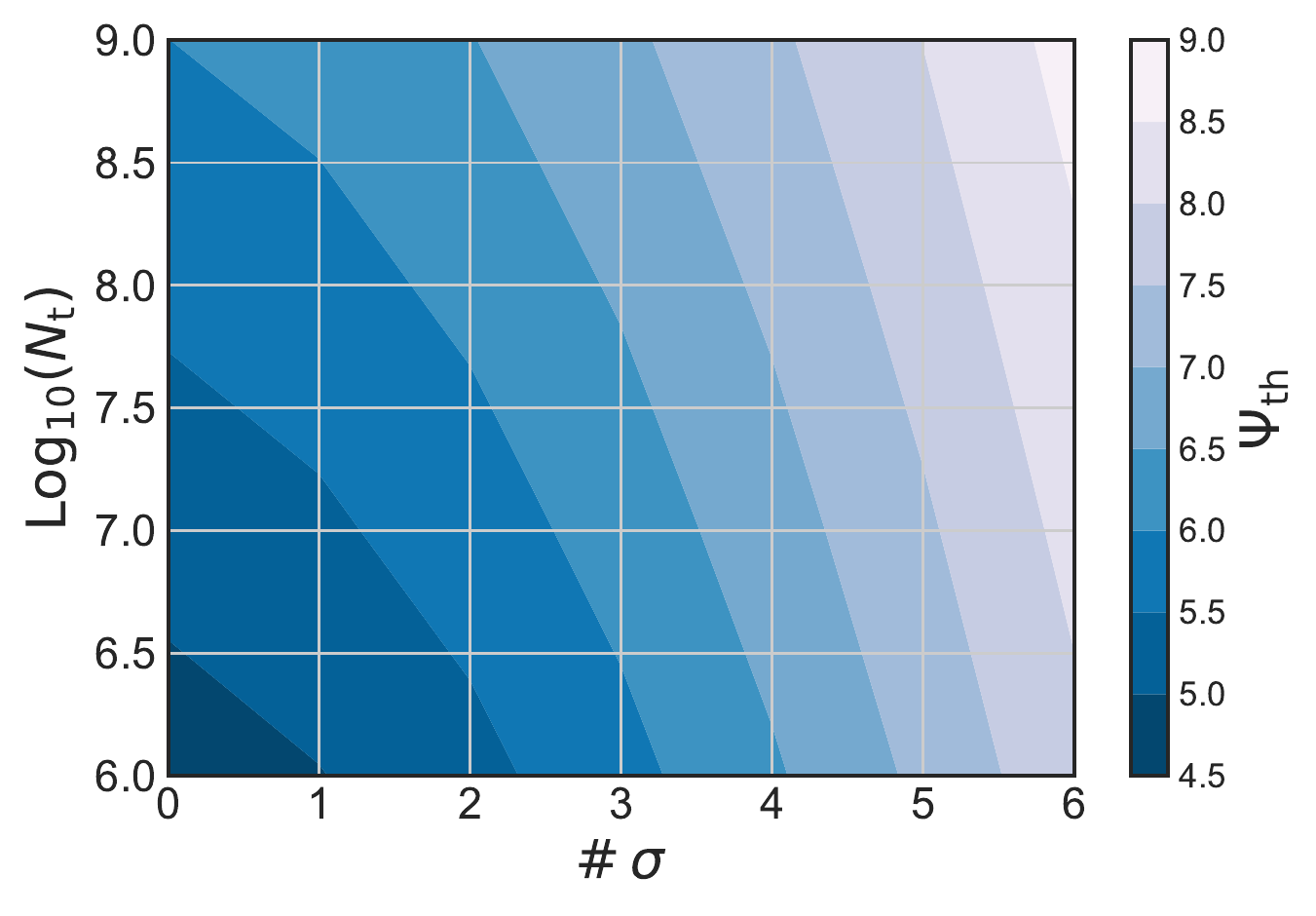}
 \vspace{-\baselineskip}
 \caption{
  \label{Ntemplates_sigmas}
  This contour plot shows how to choose a threshold $\Psi_\mathrm{th}$ for different false-alarm configurations.
  The vertical axis gives the number of templates used in a search
  and the horizontal axis shows the desired significance of candidates above threshold
  in terms of a `number of sigmas' for a Gaussian distribution.
  The color scale gives the required $\Psi_\mathrm{th}$ for candidates to reach the desired significance
  when including the trials factor from the large template bank.
 }
\end{figure}

The threshold for a search is determined under the assumption of detector noise following a stationary zero-mean Gaussian distribution
with a power spectral density $S_\mathrm{n}(f)$.
A template is considered as a candidate when its $\Psi$ exceeds a pre-specified threshold
for which the probability of a false alarm due to noise alone is small.
(See Fig.~\ref{Ntemplates_sigmas}.)
The overall false-alarm probability $\alpha_\mathrm{S}$ of the search
can be approximated as the product of the number of trials
(i.e number of templates $N_\mathrm{t}$)
and the previously introduced false-alarm probability $\alpha_\mathrm{I}$.
Now we can rewrite Eq.~\eqref{phithreshold1} in terms of the critical-ratio threshold $\Psi_\mathrm{th}$:
\begin{equation}
 \label{thresholdpsi}
 \Psi_\mathrm{th}=\sqrt{2}\,\erfc^{-1}\left(2\frac{\alpha_\mathrm{S}}{N_\mathrm{t}}\right),
\end{equation}

If the critical ratio in a template exceeds the threshold,
as a follow-up veto we can rephrase the question and consider each detector as an independent single trial,
obtaining a threshold $\Psi_\mathrm{th}^\mathrm{D}$ for each detector.
This threshold will correspond
to Eq.~\eqref{thresholdpsi} with \mbox{$N_\mathrm{t}=1$} and any given template that fails to satisfy it
in either detector
will be vetoed.

\subsection{The time-inconsistency veto} 

\begin{figure}
 \includegraphics[width=\columnwidth]{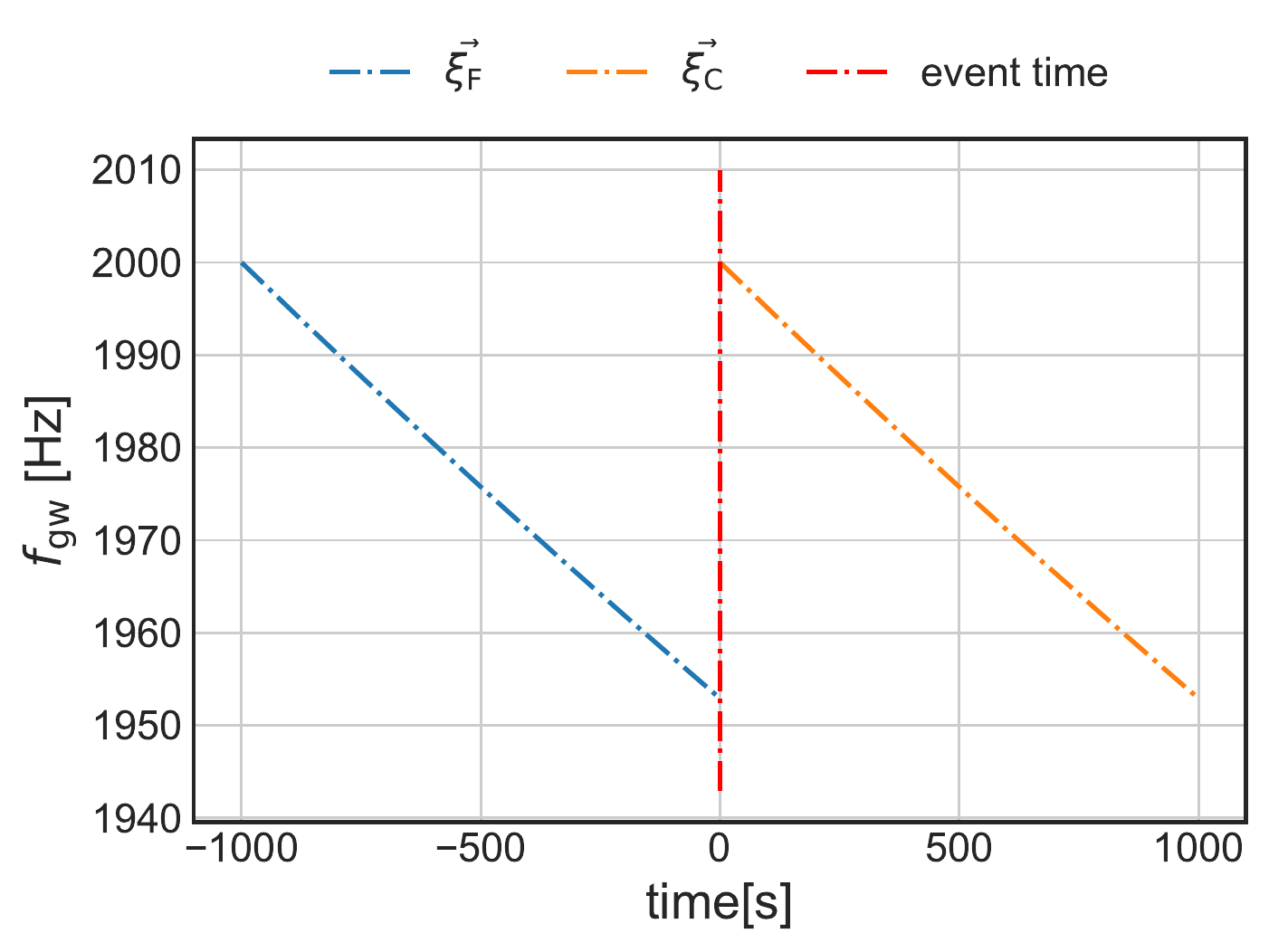}
 \vspace{-\baselineskip}
 \caption{
  For the time-inconsistency veto, we consider time-shifted frequency tracks.
  The plot shows the frequency track in time domain for a candidate template
  \mbox{$\vec{\xi_\mathrm{C}}=(f_{\mathrm{gw},0}=500$\,Hz$,n=5,\tau=10^4\,\mathrm{s},T_0=0$s)}
  and a shifted template
  \mbox{$\vec{\xi_\mathrm{F}}=(f_{\mathrm{gw},0}=500$\,Hz$, n=5, \tau=10^4\,\mathrm{s}, T_0=-T_\mathrm{obs})$},
  showing that there is no overlap between the two tracks.
  Hence, the significance $\Psi_\mathrm{F}$ of the shifted track can be used for a veto.
  }
\label{followupdist}
\end{figure}

To check that the transient behavior of the signal
matches our model, we introduce an additional veto.
Let us consider a candidate template
\mbox{$\vec{\xi_\mathrm{C}}=(f_{\mathrm{gw},0},n,\tau,T_0=T_\mathrm{event})$}
and a time-shifted version
\mbox{$\vec{\xi_\mathrm{F}}=(f_{\mathrm{gw},0},n,\tau,T_0=T_\mathrm{event}+T_\mathrm{F})$}.
These will be completely independent if
\mbox{$T_\mathrm{F} = -T_\mathrm{obs}$};
see Fig.~\ref{followupdist} for an example.
Other time shifts could be used for a veto as well,
as long as the contribution of the candidate signal $\vec{\xi_\mathrm{C}}$ to $\Psi_\mathrm{F}$
of the shifted template $\vec{\xi_\mathrm{F}}$ is zero.

The obtained value $\Psi_\mathrm{F}$ will indicate
how much of the original candidate's $\Psi_\mathrm{C}$ seems to come from a stationary contribution instead.
Stationary spectral line artifacts are common in the LIGO data \citep{Covas:2018oik} and hence this veto is important
to remove non-astrophysical false candidates.
In other words, we assign a probability to stationary lines to be the cause of the candidate.
To estimate this probability we reuse Eq.~\eqref{thresholdpsi} for a single follow-up trial.
If the resulting probability corresponds to more than 6 sigmas,
we can safely reject the candidate.

\section{Search sensitivity}

In Eq.~\eqref{sensitivityfunct} we have obtained an estimate for the sensitivity of a search
as the smallest amplitude that would cross the number-count threshold
for a given false-alarm rate $\alpha_\mathrm{I}$ and false-dismissal rate $\beta_\mathrm{I}$. 

\begin{figure}
 \includegraphics[width=\columnwidth]{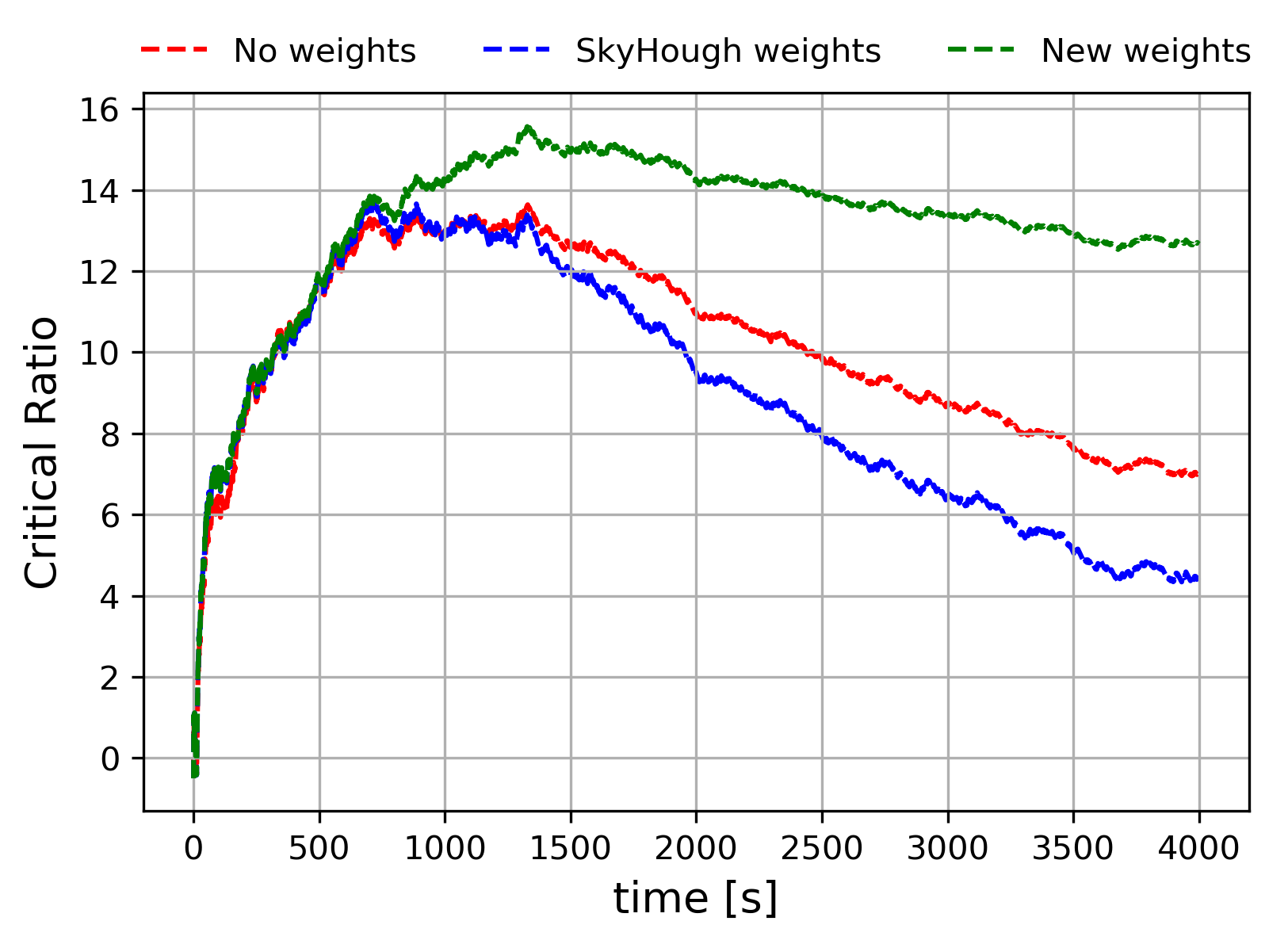}
 \vspace{-\baselineskip}
 \caption{
 \label{CRinjection}
  \updated{
  Critical ratio as a function of time using \mbox{$T_\mathrm{coh}=1$\,s} and different weights,
  for an injection at \mbox{$0.1$\,Mpc}
  in the actual aLIGO data after GW170817
  with NS parameters \mbox{$f_{gw,i} = 1565.8$\,Hz}, \mbox{$\tau = 1000$\,s}, \mbox{$n = 5$},
  \mbox{$I_\mathrm{zz}=4.34 \times 10^{38}$\,kg\,m$^2$} and \mbox{$\cos\iota=1$}
  (same as in Fig.~\ref{injection}).
  The `SkyHough weights' correspond to the scheme from \citep{WHough},
  whereas the `new weights` include source amplitude decay.
  }
 }
\end{figure}

As a specific astrophysical case, let us concentrate on
the isolated non-axisymmetric magnetar scenario as considered in the GW170817 long-duration postmerger search \citep{GW170817remnant}.
In this model, the amplitude exponent $m$ in Eq.~\eqref{amplitude-model} takes a nominal value of 2 and the signal amplitude
\updated{$h_0(t)$ is given by Eq.~\eqref{ampm2}.\footnote{\updated{In the case
of \gw[s] emitted from r-mode oscillations,
we have instead $n\lesssim7$, $m=3$ and $h_0(t)$ is given by Eq.~\eqref{amp-rmodes}.
This case is described in more detail e.g. in \cite{Owen1998}.}}
}

\updated{In Fig.~\ref{CRinjection}
we show an example signal recovery for the same injection as in Fig.~\ref{injection}.
As we can see,
power-law templates in principle allow to succesfully track this type of signal even without weights,
but including the source's amplitude decay in the weights from Sec.~\ref{sec:calibration}
is crucial for robust recovery and to fully profit from long observation times.}

Combining \updated{the amplitude from Eq.~\eqref{ampm2}} with the sensitivity as given by Eq.~\eqref{sensitivityfunct},
the astrophysical range of the search is
\begin{equation}
 \label{eq:distance}
 d = \frac{4 \pi^2 G I_\mathrm{zz} \epsilon f_{\mathrm{gw},0}^2}{c^4}\frac{\sqrt{T_\mathrm{coh}}}{3.38 S^{1/2}}\Big(\frac{\textbf{w} \cdot \textbf{X}}{||\textbf{w}||}\Big)^{(1/2)} \,.
\end{equation}

\begin{figure}
 \includegraphics[width=\columnwidth]{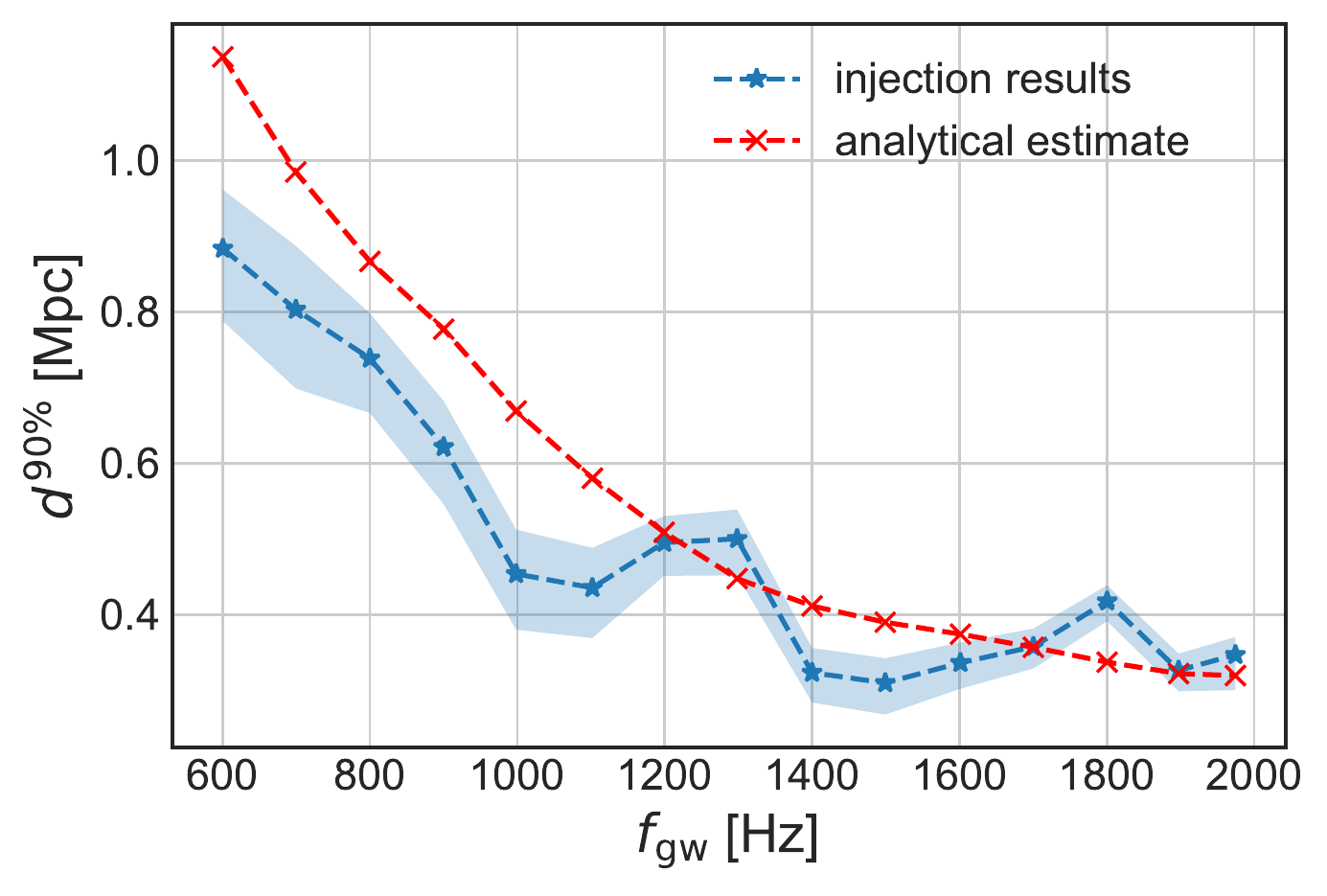}
 \vspace{-\baselineskip}
 \caption{
 \label{arange}
  A comparison of analytically and empirically obtained sensitivity estimates for a GW170817 post-merger analysis with the ATrHough method.
  The analytic sensitivity estimate was done
  for aLIGO sensitivity $S_\mathrm{n}$ during the GW170817 event (end of O2)
  and for $T_\mathrm{coh}=8$\,s.
  The empirical results correspond to the sensitive distance at 90\% detectability, $d^{90\%}$,
  obtained for the $T_\mathrm{coh}=8$\,s injection set in \cite{GW170817remnant},
  using actual aLIGO data after GW170817 and NS parameters of
  \mbox{$I_\mathrm{zz}=4.34 \times 10^{38}$\,kg\,m$^2$} and \mbox{$\cos\iota=1$},
  as well as $f_{\mathrm{gw},0}$, $\tau$ and $\epsilon$ as given in Fig.~\ref{epsmap}.
  See the appendix B of \cite{GW170817remnant} for additional results at different $T_\mathrm{coh}$.
 }
\end{figure}

\begin{figure}
 \includegraphics[width=\columnwidth]{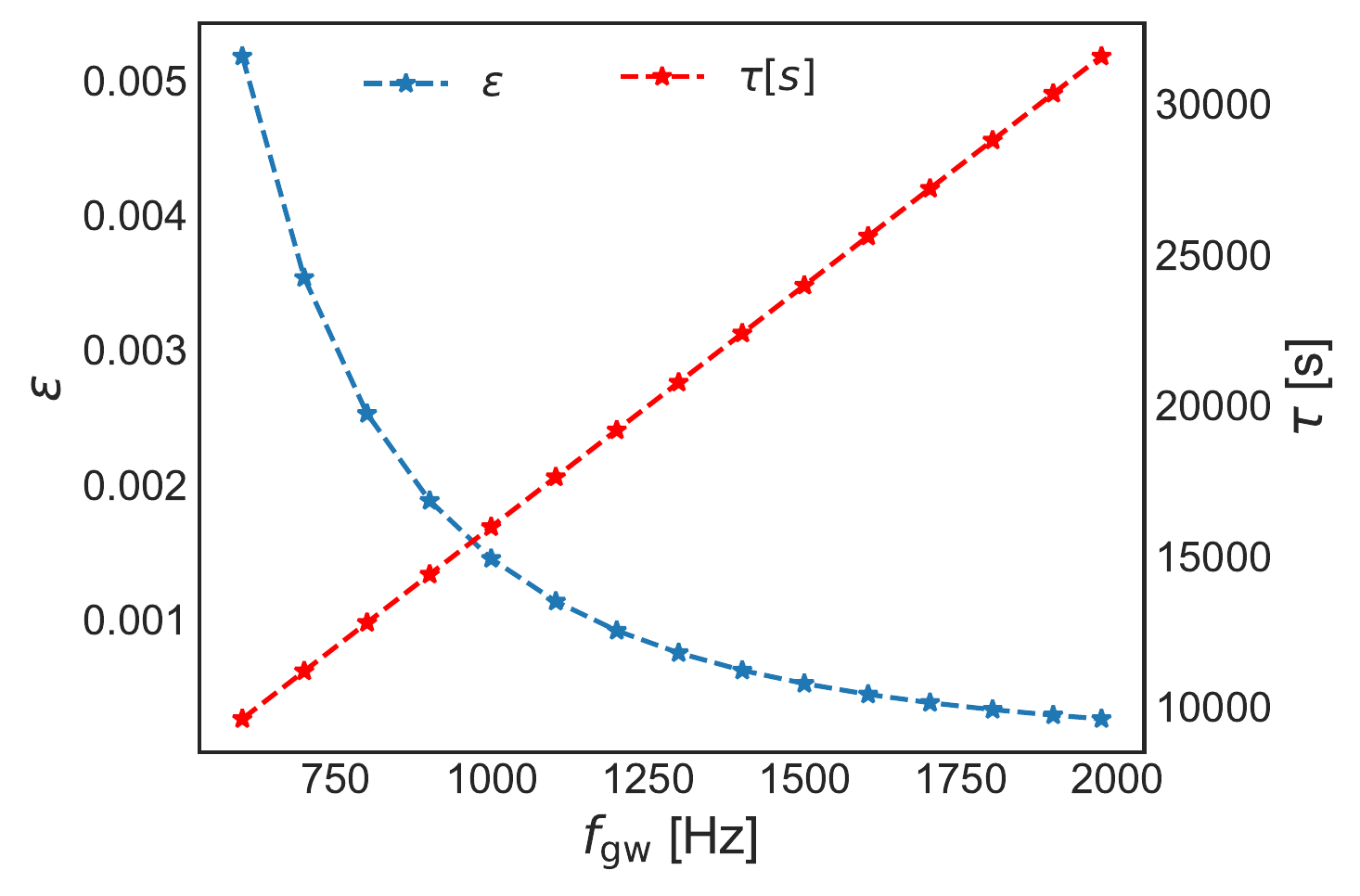}
 \vspace{-\baselineskip}
 \caption{
  \label{epsmap}
  Parameters for the $T_\mathrm{coh}=8$\,s injection set from \cite{GW170817remnant},
  as also used for the comparison with the empirical sensitivity estimate in Fig.~\ref{arange}.
  Each set of values shown corresponds to the central value of an injection subset,
  with the parameters then further randomized in narrow ranges as described below.
 }
\end{figure}

We now calculate an astrophysical range estimate for a search setup corresponding to the ATrHough analysis
performed as one of four searches in \cite{GW170817remnant}.
We use the aLIGO O2 sensitivity $S_\mathrm{n}$ during the GW170817 event to calculate the weights,
and for the remnant's moment of inertia we use the same value as in \cite{GW170817remnant},
\mbox{$I_\mathrm{zz}=100 M_\odot^3 G^2/c^4\approx 4.34 \times 10^{38}\,\mathrm{kg}\,\mathrm{m}^2$}.

In Fig.~\ref{arange} we compare the analytical estimate with the empirical recovery fraction for a set of injections.
Those were originally performed for the sensitivity estimate in the GW170817 post-merger search \citep{GW170817remnant}.
The recovery criterion corresponds to \mbox{$\Psi_\mathrm{th}=9$},
or a $5\sigma$ significance.
We have concentrated here on a braking index \mbox{$n = 5$} that corresponds to pure \gw emission,
and covered ranges of $f_{\mathrm{gw},0}$ and $\tau$ as shown in Fig.~\ref{epsmap}.
The procedure to obtain the experimental results consisted in selecting 10\,Hz wide frequency bands,
for each band injecting 1000 simulated signals into O2 data with amplitudes around the astrophysical range estimate.
The purpose was to find the amplitude corresponding to $90\%$ recovery efficiency.
The parameters $\tau$ and $f_{\mathrm{gw},0}$ were randomized within 10 bins of their nominal value;
i.e. the injection parameters are not perfectly aligned with the search grid,
thus allowing for a realistic exploration of search mismatch in the recovery.

We do not expect an exact agreement between analytical prediction and sensitivity measured from injections,
as the analytical estimate is based on a Gaussian noise approximation.
But the results are sufficiently close to demonstrate that Eq.~\eqref{sensitivityfunct} is useful
for the purpose of setting up future searches.

\section{CONCLUSIONS}

In this paper we have described a new semi-coherent search method
\updated{for quasi-monochromatic gravitational waves},
using short incoherent steps of the order of seconds
with the intention to track signals of intermediate durations (of the order of hours to days)
\updated{even if these show rapid frequency evolution}.
\updated{The main innovations compared to previous versions of the Hough transform method~\citep{WHough,Sintes2006,Hough}
are the new frequency-evolution templates
and the additional inclusion of amplitude evolution in the Hough weights.}

\updated{In introducing this new method
and estimating its sensitivity,}
we have concentrated on the model of power-law spin-down for a newborn \ns.
As applied in the GW170817 post-merger remnant search \citep{GW170817remnant},
the astrophysical range of this method at 90\% detection confidence is at $\sim 1$\,Mpc
with LIGO sensitivity at the end of the second observing run.
With future instruments like the Einstein Telescope \citep{Punturo:2010zza,Hild:2010id,Sathyaprakash:2012jk},
this range could increase by a factor of $\sim 20$.

One disadvantage of modeled semi-coherent methods like this one
is the need to explicitly set a starting time for the signal model.
On the other hand,
it is a suitable method to perform fast and economic follow-ups of known merger events
or for promising candidates identified by more generic searches,
allowing to reliably set up a fixed false-alarm rate of the overall search.

The same strategy can also easily be translated to signals following other spin-down patterns
than the power-law model we focused on so far,
with the definition of weights and parameter space grids following the same procedure as introduced in this paper.

\section*{Acknowledgments}

We thank the LIGO-Virgo Continuous Wave working group and the GW170817 postmerger search team,
in particular
S.~Banagiri,
M.~Bejger,
A.~Miller,
L.~Sun,
K.~Wette
and S.~Zhu,
for many fruitful discussions.
M.O. and A.M.S. acknowledge the support of the Spanish
Agencia Estatal de Investigaci\'on
and Ministerio de Ciencia, Innovaci\'on y Universidades
grants FPA2016-76821-P, FPA2017-90687-REDC, FPA2017-90566-REDC, and FPA2015-68783-REDT,
the Vicepresidencia i Conselleria d'Innovaci\'o, Recerca i Turisme del Govern de les Illes Balears,
the European Union FEDER funds,
and the EU COST actions CA16104, CA16214 and CA17137.
The authors are grateful for computational resources provided by the 
LIGO Laboratory and supported by National Science Foundation Grants PHY-0757058 and PHY-0823459.

\appendix
\section{Testing the Gaussian approximation for the weighted number count $\nu$}
\label{sec:gaussiankl}

\newupdated{In Eq.~\eqref{gaussianapprox}
we have approximated the distribution $p(\nu|\rho_\mathrm{th},\lambda)$
of the weighted number-count statistic $\nu$,
when using appropriate weights and for a sufficient number of SFTs,
by a Gaussian.
Here we present some simple empirical tests of this limiting behaviour
in configurations similar to the search implemented in~\cite{GW170817remnant}.
}

\begin{figure}
\begin{center}
\includegraphics[width=0.4\textwidth]{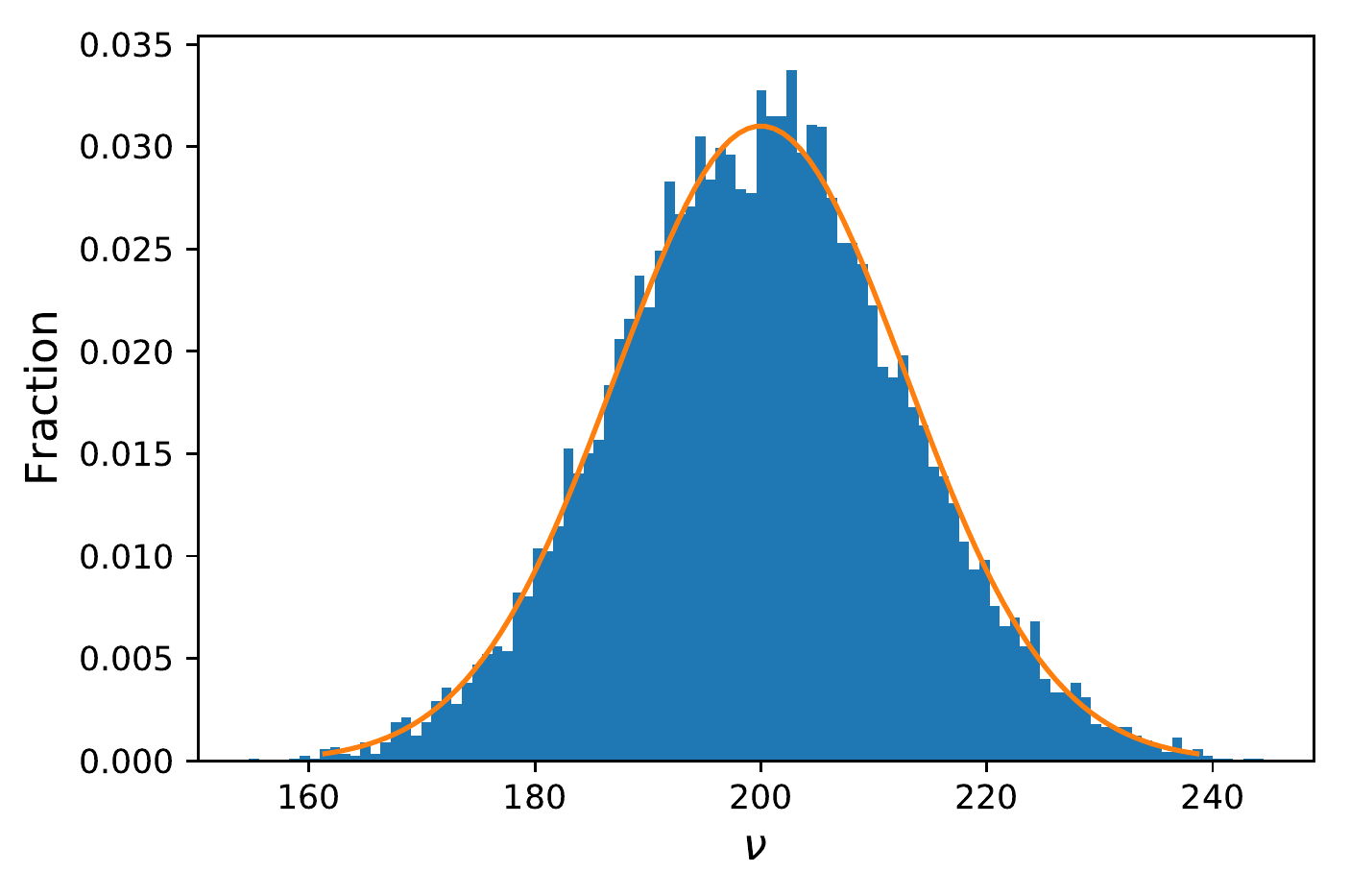}
\caption{
 \newupdated{Example of the close agreement
 between empirical $\nu$ results in pure Gaussian noise
 and the Gaussian approximation from Eq.~\eqref{gaussianapprox}.
 Over 10000 templates,
 this example yields a KL divergence of $\approx3\times10^{-5}$.
 }
}
\label{figure1}
\end{center}
\end{figure}

\begin{figure}
\begin{center}
\includegraphics[width=0.4\textwidth]{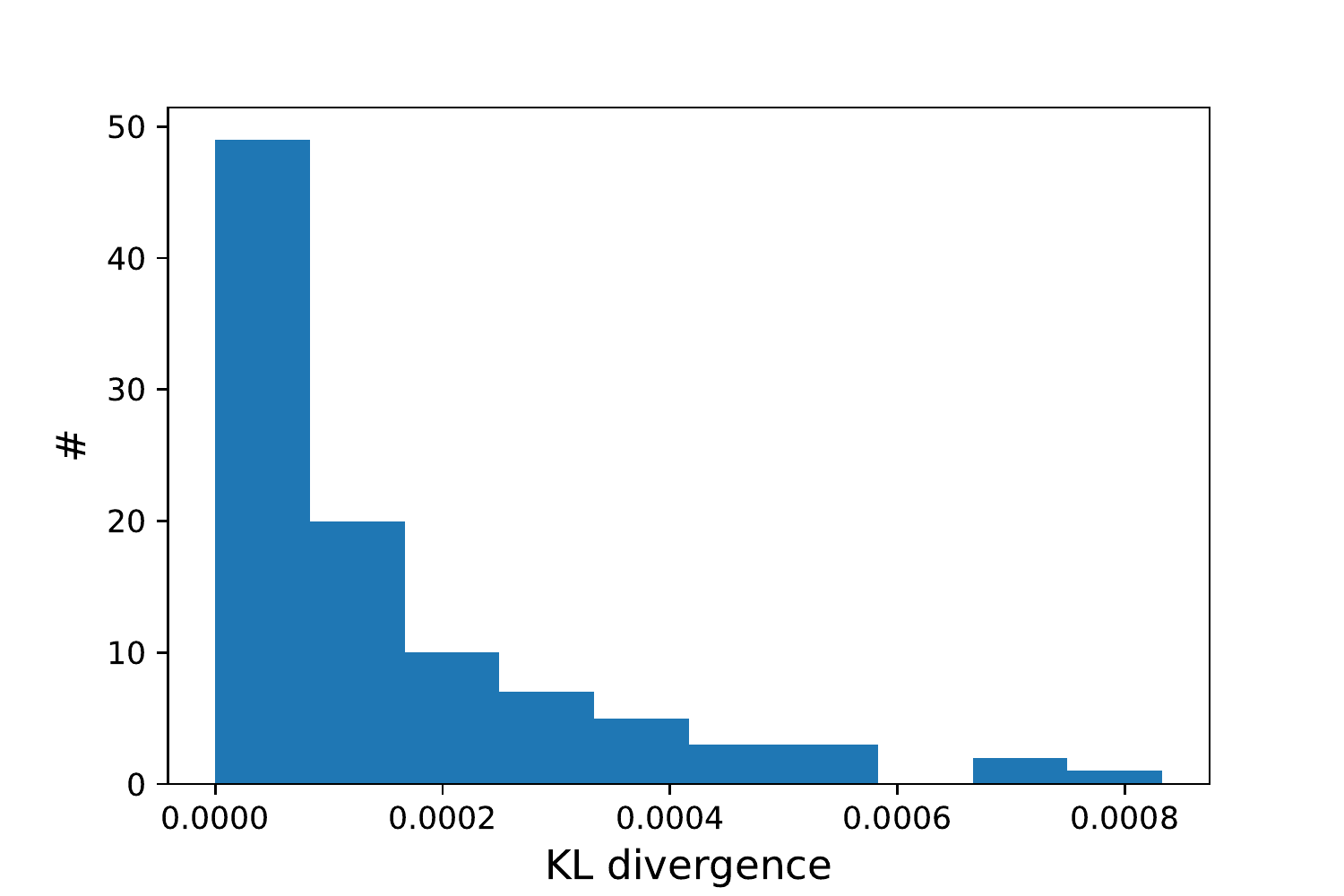}
\caption{
 \newupdated{Histogram of KL divergences for 100 simulations
 \mbox{($h_0 = 0$)} on 1000 segments of Gaussian noise each,
 analysed with 10000
 templates.
 }
}
\label{figure2}
\end{center}
\end{figure}

\newupdated{Using the same machinery as before,
we have analysed 100 simulated data sets,
each consisting of 1000 segments of Gaussian noise with no GW injection (\mbox{$h_0 = 0$}).
For each, we have computed the number count for 10000 template trials,
covering a small fraction of the parameter space around a random point corresponding to the `null injection',
and using the weights proportional to $X_i$ as introduced in Sec.~\ref{sec:calibration}.
We have then compared the resulting empirical distribution of $\nu$ with our Gaussian approximation from Eq.~\eqref{gaussianapprox}.
An example is shown in Fig.~\ref{figure1} to illustrate the agreement between the two distributions.
}

\newupdated{To further evaluate the (dis-)agreement between two distributions $P$ and $Q$,
one can compute the Kullback-Leibler (KL) divergence~\citep{kullback1951}
(in bits):
\begin{equation}
 \label{KLdiver}
 D_\mathrm{KL}(P \parallel Q) = \sum_{x\in\mathcal{X}} P(x) \log_2\left(\frac{P(x)}{Q(x)}\right) \,,
\end{equation}
for a discrete set $\mathcal{X}$ of measured values.
Note the asymmetry in this definition;
here we take the Gaussian for $P$
and the empirical results for $Q$.
A histogram of KL divergences between
the Gaussian from Eq.~\eqref{gaussianapprox}
and the empirical distributions from the 1000 simulations
is shown in Fig.~\ref{figure2}.
We see that there is far less than 1 bit of information between the two distributions
in all draws.
Hence, based on the KL divergence we can consider the approximation from Eq.~\eqref{gaussianapprox}
as a sufficiently robust basis for estimating significance of our search results.
}


\bibliography{references.bib}


\end{document}